\documentclass[12pt]{article}
\usepackage{lscape}
\newcommand{\ft}[2]{{\textstyle\frac{#1}{#2}}}
\newsavebox{\zzzbar}
\sbox{\zzzbar}
  {\setlength{\unitlength}{0.9em}
  \begin{picture}(0.6,0.7)
  \thinlines
  \put(0,0){\line(1,0){0.6}}
  \put(0,0.75){\line(1,0){0.575}}
  \multiput(0,0)(0.0125,0.025){30}{\rule{0.3pt}{0.3pt}}
  \multiput(0.2,0)(0.0125,0.025){30}{\rule{0.3pt}{0.3pt}}
  \put(0,0.75){\line(0,-1){0.15}}
  \put(0.015,0.75){\line(0,-1){0.1}}
  \put(0.03,0.75){\line(0,-1){0.075}}
  \put(0.045,0.75){\line(0,-1){0.05}}
  \put(0.05,0.75){\line(0,-1){0.025}}
  \put(0.6,0){\line(0,1){0.15}}
  \put(0.585,0){\line(0,1){0.1}}
  \put(0.57,0){\line(0,1){0.075}}
  \put(0.555,0){\line(0,1){0.05}}
  \put(0.55,0){\line(0,1){0.025}}
  \end{picture}}
\newcommand{\Zbar}{\mathord{\!{\usebox{\zzzbar}}}}
\newsavebox{\uuunit}
\sbox{\uuunit}
    {\setlength{\unitlength}{0.825em}
     \begin{picture}(0.6,0.7)
        \thinlines
        \put(0,0){\line(1,0){0.5}}
        \put(0.15,0){\line(0,1){0.7}}
        \put(0.35,0){\line(0,1){0.8}}
       \multiput(0.3,0.8)(-0.04,-0.02){12}{\rule{0.5pt}{0.5pt}}
     \end {picture}}
\newcommand {\unity}{\mathord{\!\usebox{\uuunit}}}

\newcommand{\OS}{$OSp(1|32)$}
\newcommand{\Poin}{Poincar\'e}
\newcommand{\rmi}{{\rm i}}
\newcommand{\Gamnul}{\lower-.2ex\hbox{$\stackrel{0}{\Gamma }$}}
\newcommand{\Znul}{\lower-.2ex\hbox{$\stackrel{0}{Z }$}}

\csname @addtoreset\endcsname{equation}{section}
\newcommand{\ncc}{\otimes}
\newcommand{\eqs}{} 
\newsavebox{\nwsea}
\sbox{\nwsea}
    {\setlength{\unitlength}{0.825em}
     \begin{picture}(3,3)
        \put(3,0){\vector(-3,2){3}}
        \put(0,2){\vector(3,-2){3}}
        \put(2,1){\mbox{$\textrm{T}_{st}$}}
     \end {picture}}
\newcommand{\nwsearrow}{\mathord{\!\usebox{\nwsea}}}
\newsavebox{\neswa}
\sbox{\neswa}
    {\setlength{\unitlength}{0.825em}
     \begin{picture}(3,3)
        \put(0,0){\vector(3,2){3}}
        \put(3,2){\vector(-3,-2){3}}
        \put(2.6,0.6){\mbox{$\textrm{T}_{st}$ }}
     \end {picture}}
\newcommand{\neswarrow}{\mathord{\!\usebox{\neswa}}}
\newsavebox{\nsa}
\sbox{\nsa}
    {\setlength{\unitlength}{0.825em}
     \begin{picture}(3,3)
        \put(1.9,0){\vector(0,1){2}}
        \put(1.9,2){\vector(0,-1){2}}
        \put(2.4,1){\mbox{$\textrm{T}_{tt}$}}
        \put(0,1){\mbox{$\textrm{T}_{ss}$}}
     \end {picture}}
\newcommand{\nsarrow}{\mathord{\!\usebox{\nsa}}}
\newsavebox{\nsass}
\sbox{\nsass}
    {\setlength{\unitlength}{0.825em}
     \begin{picture}(3,2.4)
        \put(1.9,0){\vector(0,1){2}}
        \put(1.9,2){\vector(0,-1){2}}
        \put(2.4,0.6){\mbox{$\textrm{T}_{ss}$}}
     \end {picture}}
\newcommand{\nsarrowss}{\mathord{\!\usebox{\nsass}}}
\newsavebox{\crtt}
\sbox{\crtt}
    {\setlength{\unitlength}{0.825em}
     \begin{picture}(3,2.4)
        \put(0,0){\vector(3,2){3}}
        \put(3,2){\vector(-3,-2){3}}
        \put(0,2){\vector(3,-2){3}}
        \put(3,0){\vector(-3,2){3}}
        \put(2.2,0.6){\mbox{$\textrm{T}_{tt}$}}
     \end {picture}}
\newcommand{\crosstt}{\mathord{\!\usebox{\crtt}}}
\newsavebox{\moei}
\sbox{\moei}
    {\setlength{\unitlength}{0.825em}
     \begin{picture}(21,6)
        \put(17,3){\vector(-1,0){13}}
        \put(17,3){\vector(4,3){2}}
        \put(9,4.5){\mbox{$\textrm{IIB}_{9,1}$}}
       \put(0,3){\mbox{$\textrm{IIA}'_{10,0}$}}
\put(1,1){\vector(0,1){1.5}}
\put(1,2.5){\vector(0,-1){1.5}}
\put(1.3,1.5){\mbox{\textrm{S}}}
 \put(0,0){\mbox{$\textrm{IIA}_{10,0}$}}
 \put(19,4.5){\mbox{$\textrm{IIB}^*_{9,1}$}}
\put(20,1){\vector(0,1){2.2}}
\put(20,3.2){\vector(0,-1){2.2}}
\put(20.3,1.7){\mbox{\textrm{S}}}
 \put(19,0){\mbox{$\textrm{IIB}'_{9,1}$}}
 \put(4,1){\vector(1,0){15}}
 \put(19,1){\vector(-1,0){15}}
 \put(11,2){\mbox{$\textrm{T}_{st}$}}
     \end {picture}}
\newcommand{\moeilijk}{\mathord{\!\usebox{\moei}}}
\newsavebox{\wegr}
\sbox{\wegr}
    {\setlength{\unitlength}{0.825em}
     \begin{picture}(5,6)
 \put(0,1){\vector(1,0){5}}
 \put(5,1){\vector(-1,0){5}}
 \put(0,4.5){\vector(1,0){5}}
 \put(5,4.5){\vector(-1,0){5}}
 \put(3,2){\mbox{$\textrm{T}_{st}$}}
     \end {picture}}
\newcommand{\weg}{\mathord{\!\usebox{\wegr}}}
\newsavebox{\nsS}
\sbox{\nsS}
    {\setlength{\unitlength}{0.825em}
     \begin{picture}(3,3)
        \put(1.9,0){\vector(0,1){2}}
        \put(1.9,2){\vector(0,-1){2}}
        \put(2.4,1){\mbox{\textrm{S}}}
     \end {picture}}

\begin{document}

\begin{titlepage}
\begin{flushright}
UG-00-03\\
KUL-TF-2000/11\\
hep-th/0003261
\end{flushright}
\vspace{.5cm}
\begin{center}
\baselineskip=16pt
{\LARGE    The many faces of $OSp(1|32)$
}\\
\vfill
{\large Eric Bergshoeff $^{1}$ and
Antoine Van Proeyen $^{2,\dagger }$ 
  } \\
\vfill
{\small
$^1$ Institute for Theoretical Physics, Nijenborgh 4,
9747 AG Groningen, The Netherlands \\[3mm]
$^2$  Instituut voor Theoretische Fysica, Katholieke Universiteit
Leuven, Celestijnenlaan 200D, B-3001 Leuven, Belgium
}
\end{center}
\vfill
\begin{center}
{\bf Abstract}
\end{center}
We show that the complete superalgebra of symmetries, including central
charges, that underlies F-theories,
M-theories and type II string theories
in dimensions 12, 11 and 10 of various signatures correspond to
rewriting of the same $OSp(1|32)$ algebra in different covariant ways.
One only has to
distinguish the complex and the unique real algebra.
We develop a common framework to discuss all signatures theories by
starting from the complex form of $OSp(1|32)$. Theories are
distinguished by the choice of basis for this algebra. We formulate
dimensional reductions and dualities as changes of basis of the algebra.
A second ingredient is the choice of a real form corresponding to a
specific signature. The
existence of the real form of the
algebra selects preferred spacetime signatures.
In particular, we show how the real $d=10$ IIA and IIB
superalgebras for various signatures are related
by generalized T-duality transformations that not only involve
spacelike but also timelike directions.
A third essential ingredient is that the translation generator in one
theory plays the role of a central charge operator in the other theory.
The identification of the translation generator
in these algebras leads to the star algebras of Hull, which are
characterized by the fact that the positive definite energy operator
is not part of the translation generators.
We apply our results to discuss different T-dual pictures of the
D-instanton solution of Euclidean IIB supergravity.
\vspace{2mm} \vfill \hrule width 3.cm
{\footnotesize
\noindent $^\dagger$  Onderzoeksdirecteur FWO, Belgium. }
\end{titlepage}
\section{Introduction}
In some recent applications of string theory an important role is
played by the D-instanton, as introduced in
\cite{Polchinski:1994fq,Green:1995my}. The brane
solution corresponding to such a D-instanton was considered in
\cite{Gibbons:1996vg}. There, the
instanton was presented as a brane solution with  transverse
directions only of a 10-dimensional Euclidean theory containing
the metric, the dilaton and the Ramond--Ramond scalar. It was
suggested that this was part of an Euclidean IIB supergravity theory
in 10 dimensions. At first sight, the definition of such an Euclidean
IIB supergravity theory seems problematic, since one cannot extend the
self-duality
constraint of the real IIB Ramond-Ramond 5-form field strength from a
Minkowski spacetime to a spacetime of signature
$(10,0)$\footnote{See the imaginary factor in (\ref{dualF}) for $d=10$,
$t=0$.}. This paper addresses the issue in which sense the D-instanton
can be represented as a brane solution of an Euclidean supersymmetric theory.

Euclidean supersymmetric theories have been considered in lower
dimensions \cite{Zum77,Nic78,Nie96}. In some cases, the Euclidean
theory could be obtained only by complexifying the fields, and considering
holomorphic actions. This procedure seems also appropriate for the
IIB Euclidean theory mentioned above, as the self-duality of the
5-form can be maintained for complex fields only\footnote{We thank
Peter van Nieuwenhuizen for a discussion on this point.}.

The D-instanton is a solution
of the Euclidean action \cite{Gibbons:1996vg}
\begin{equation}
S_{\rm E} = \int d^{10}x \sqrt{g}\left[ R  - \ft12(\partial \phi)^2
+ \ft12 e^{2 \phi} (\partial \ell)^2
\right]  \,,
\label{Euclaction}
\end{equation}
where $\phi $ is the dilaton and $\ell $ is the Ramond--Ramond
scalar. In the solution, the metric is flat, $g_{\mu \nu }=\delta _{\mu \nu }$, and
\begin{equation}
\pm \ell = e^{-\phi} = H^{-1}\,,
\end{equation}
where $H$ is a general harmonic function over
the 10-dimensional flat Euclidean space. If we want to connect the
D-instanton with other branes by T-duality, we have to reduce it to
a spacetime of signature
$(9,0)$ thus yielding a D-instanton in 9 Euclidean dimension. The
only standard D-brane that also gives rise, upon reduction,
to a D-instanton in 9
Euclidean dimensions is the D0-brane of the type IIA string. This D0-brane
is a solution of the Minkowskian action
\begin{equation}
  S_{\rm M} = \int d^{10}x \sqrt{-g}\left[- R  - \ft12(\partial \phi)^2
  -\ft14 e^{3\phi /2}F_{\mu \nu }F^{\mu \nu }\right]  \,,
\label{Minkaction}
\end{equation}
given by
\begin{eqnarray}
ds^2 & = & -H^{-7/8}dt^2+H^{1/8}dx^2_{(9)}\, , \nonumber\\
e^\phi  & = & H^{3/4}\,,\qquad A_t=\pm H^{-1}\,,
\label{solD0}
\end{eqnarray}
where $H$ is an harmonic function over the 9-dimensional Euclidean
transverse space.
The D-instanton and the D0-brane are mapped to each other
under the type II T-duality rules \cite{Bergshoeff:1995as}
adapted to the Euclidean case for the D-instanton, as follows (we use
Einstein frame) \cite{Bergshoeff:1998ry}
\begin{equation}
  \begin{array}{ccl}
  {\rm IIB,\ (10,0)} & \leftrightarrow & \hskip .8truecm {\rm IIA,\ (9,1)} \\ \hline && \\[-3mm]
g_{\mu \nu } & \leftrightarrow &\hskip .4truecm e^{\phi /8}g_{\mu \nu }
(-g_{tt})^{1/4}\, ,\\
    g_{xx} & \leftrightarrow &\hskip .4truecm e^{-7\phi /8}(-g_{tt})^{-3/4}\, , \\
    \ell  & \leftrightarrow &\hskip .4truecm A_t\, , \\
    e^\phi  & \leftrightarrow &\hskip .4truecm e^{3\phi /4}(-g_{tt})^{-1/2} \,,
  \end{array}
\label{dualityrules}
\end{equation}
where $x$ in the IIB and $t$ in the IIA case are the directions in which
the duality is
performed, and $\mu ,\nu $ are the remaining directions.

To obtain the above duality transformation, we had to
compactify one theory over a spacelike direction, and the other over
a timelike direction\footnote{The case in which both theories are
compactified over a timelike direction has been studied in
\cite{Moore,Cremmer:1998em}.
Relations between Euclidean and Minkowski supersymmetry
by reducing over a different number of space versus time directions played
an important role in \cite{Belitsky:2000ii}.}.
 This has to be contrasted with the usual
duality transformations, where both theories are compactified over a
spacelike direction. To distinguish these dualities, we will refer to
the usual type of duality as a space/space duality, and the above one
as a space/time duality. The special feature of a space/time duality
is that it connects theories with different signature of spacetime.
Ten- and eleven-dimensional theories with various signatures, connected by
different space/time dualities have been studied in \cite{Hull}.
For further work on these theories, see \cite{StarTheories}. The
purpose of this paper is to reconsider the results of \cite{Hull}
from a superalgebra point of view\footnote{Some remarks on the
underlying (complex)  algebraic structure were already made in \cite{Hull}.}.
All the theories that we consider
have 32 supersymmetries\footnote{For earlier  work on superalgebras
with 32 supersymmetries, see \cite{Bars:1997ab}.
Superalgebras with more than
32 supersymmetries, containing both IIA and IIB algebras, have been
considered in e.g.~\cite{Bars:1997ug,Rudychev:1998ui}.}.
Including all the central charges, the
superalgebra must then be (a contraction of) \OS, which is
the {\it unique} maximal simple
superalgebra with 32 fermionic generators. We will show in which
sense all the
theories occurring in \cite{Hull}
 are indeed related by this common algebraic structure. In
the \OS\ algebra, the bosonic generators that occur at the right-hand
side of the $\{Q,Q\}$ anticommutator have non-trivial commutation
relations, both between themselves and with the supersymmetry
generators. In
the \Poin\ theories, which we consider, the algebra is in fact the
contraction of the \OS\ algebra such that all these commutators
vanish. One might also consider anti-de Sitter--type theories where
the uncontracted \OS\ algebra exists. The connection to anti-de
Sitter has been made in \cite{vanHolten:1982mx}, by identifying the
2-index generator at the right-hand side of $\{Q,Q\}$ with the
Lorentz generator $M_{\mu \nu }$, such that
$[P_\mu ,P_\nu ]=M_{\mu \nu }$.

The maximal dimension in which the generators of \OS\ can be
considered as representations of a Lorentz algebra is twelve. In the
decomposition of the generators of \OS\ into representations of the
12-dimensional Lorentz algebra there is no vector, i.e.\ there is no
room for translations. Translations are essential to define a
spacetime, and thus a supersymmetric action. Only after reduction to
11 dimensions translations appear in the algebra, and we can start to
consider actions. We will see that in various ways of reduction
(first over spacelike or timelike directions), the translations get
identified with different generators of the 11-dimensional algebra.
To identify the symmetry algebras of  the different theories
occurring in \cite{Hull},
generators that play the role of translations in one theory, become
central charges in others. The appearance of different twisted algebras in
\cite{Hull} can be understood from this point of view, and leads in
our approach to identical complete superalgebras once the central
charges have been included.

As mentioned above, sometimes it is necessary to work with complex fields.
This reflects itself in the necessity to use the complex form of \OS.
Only for certain signatures one can use the real form of the algebra and
work with real fields in the corresponding supergra\-vity theory.
We remind the reader that the complex form of an algebra
with generators $T_A$ means that the algebra vector space consists of
all elements $ \epsilon ^A T_A$ with $\epsilon ^A$ complex numbers.
The real form is a subspace of this vector space defined by a reality
condition on all the $\epsilon ^A$, e.g.\ that they are all real. In
general, a complex form of a (super-) algebra can have different real
forms. Consider for instance the 3 generators of $SU(2)$ with commutation
relations
\begin{equation}
  [T_i, T_j]=\varepsilon _{ijk}T_k\, .
\label{SU2generators}
\end{equation}
The real form defined by $a^iT_i$ with all $a^i$ real is the real form of
$SU(2)$, while $\rmi b^1 T_1 +\rmi b^2 T_2+ b^3T_3$ with all $b^i$
real defines $SU(1,1)$. Conveniently rewriting the above expression as $b^iS_i$,
such that the factors of $\rmi$ disappear, the commutation relations are
\begin{equation}
  [S_1,S_2]=-S_3\,,\qquad [S_2,S_3]=S_1\,,\qquad [S_3,S_1]=S_2\,,
\label{SU11generators}
\end{equation}
showing a sign difference in the commutation relations. Note that this sign
is not relevant in the complex form, as $b^i S_i$ with all $b_i$
complex is the same complex algebra as the one mentioned in the
$T$-basis.

An important property is that \emph{\OS\ has only one real form}. This
can be seen by looking at the list of real forms of superalgebras
established in \cite{realLieSA} and conveniently summarized in a
table in \cite{Claus:1998us,VanProeyen:1999ni}. In order that this
real form can be used for a particular dimension and signature, the
parameters should form a real 32-dimensional representation of the
corresponding Lorentz algebra. We can thus determine the relevant
cases by considering for which signatures the smallest spinor has
real dimension at most 32. Spinors for arbitrary spacetime signatures
have been studied in \cite{Gammamatrices}
and for a convenient
table, one can consider table~2 in
\cite{VanProeyen:1999ni}. This shows that the highest dimension is
again 12, but only for signatures $(10,2)$, $(6,6)$ or $(2,10)$. In
11 dimensions the allowed signatures are $(10,1)$, $(9,2)$, $(6,5)$
or the $s\leftrightarrow t$ interchanged ones (which are
equivalent using
redefinitions). In 10 dimensions and lower, all signatures are
possible, sometimes with irreducible spinors and otherwise as type II
theories (see Table~\ref{tbl:MWSspinors} in Section~\ref{ss:realalgebras}).
The complex form of the algebra is independent of the signature of
spacetime. Indeed, the complex form of the Lorentz algebra for any
signature is the same, similar as we saw above that the complex form
of $SU(2)$ is the same as that of $SU(1,1)$.

We will show in this paper that manipulations such as
reductions and dualities can be understood as different
reparametrizations of the same \OS\ algebra, in its complex or real form.
\emph{Thus all different possibilities of M-type or type  II theories
can be viewed as different faces of the same \OS\ superalgebra},
explaining the title of this paper.\vspace{5mm}

In section~\ref{ss:complexalgebras}, we will consider the complex
form of \OS, written as a symmetry algebra in dimensions ten,
eleven and twelve. It
is convenient to start with the complex form, as this is independent
of the signature. We will discuss the formulation of the algebras in
these dimensions, their dimensional reductions and the (complex)
T-dualities relating them.
Section~\ref{ss:realalgebras} considers the restriction of these
complex algebras to real algebras, for which the signature becomes
important. In parallel to the previous section we will first discuss
the formulation of the real algebras in the different dimensions
and next their dimensional reductions and the (real) T-dualities
relating them. The final result is summarized in Table~\ref{ff:small}.
In Section~\ref{ss:SUGRA} we discuss the relation between
the superalgebra and the corresponding supergravity action.
In particular, we will identify
the translations between the bosonic charges,
and show that different choices of the translation
generator (interchanging it with central charges) lead to the different $\star$-algebras of \cite{Hull}.
The different theories thus obtained are
summarized in Table~\ref{ff:detail} which is similar to
the one given in \cite{Hull}. Finally, in
the Conclusions  we will come back to the issue of the
D-instanton and its T-dual formulations, as discussed in the
beginning of this Introduction.

\section{Complex symmetry algebras}\label{ss:complexalgebras}
\subsection{Algebras in  12 to 10 dimensions}\label{ss:calg1210}
We consider algebras with 32 supersymmetries, not taking into account
their reality properties. This implies that $d=12$ is the highest
dimension. In this subsection, we present the algebras in 12 to 10
dimensions, leaving the relations between them to the next two subsections.
We start by writing the \OS\ algebra in a
12-dimensional covariant way. This means that the anticommutator of
the supersymmetries is the most general expression in accordance with
the symmetries, i.e.\ it reflects the decomposition of the
symmetric product of two 32-dimensional spinor representations of $Sp(32)$ in $SO(12)$
representations\footnote{In this section we discuss the complex algebras,
and thus the notation $SO(12)$ should be understood as the
corresponding algebra over the complex field.}:
\begin{equation}
(32\times 32)_S= 66 + 462^+\,.
\label{numbers12}
\end{equation}
The 32 spinor components are a chiral spinor in 12 dimensions, i.e.\
$ \hat{\Gamma }_*\hat{Q}=\hat{Q}$. For the notations of spinors,
gamma matrices, and duals of tensors, we refer to appendix~\ref{app:conventions}.
Note that for the complex case, the signature of spacetime is not
relevant. However, we have to choose a metric for raising and
lowering indices, \ldots~. We will use the notations such that this
metric is the identity, which implies that we can use the general
formula as if we are in the Euclidean case $t=0$.
The anticommutator of two supersymmetries is\footnote{Observe the omission
of the charge conjugation matrix in our notation, see (\ref{shortanticomm}).}
\begin{equation}
\left\{\hat{Q},\hat{Q} \right\} =\ft12 \hat{{\cal P}}^+ \hat{\Gamma
}^{\hat{M}\hat{N}}\hat{Z}_{\hat{M}\hat{N}} +\ft1{6!} \hat{{\cal P}}^+ \hat{\Gamma
}^{\hat{M}_1\cdots \hat{M}_6}\hat{Z}^+_{\hat{M}_1\cdots \hat{M}_6}\,,
\label{calg12}
\end{equation}
where the gamma matrices are $64\times 64$ matrices, but due to the
chiral projection operators $\hat{ {\cal P}}{}^+$, their relevant part is $32\times
32$.
The 6-index generator is selfdual, i.e.\
\begin{equation}
  \hat{Z}^+_{\hat{M}_1\cdots \hat{M}_6}=
\tilde{\hat{Z}}{}^+_{\hat{M}_1\cdots \hat{M}_6}\,,
\label{selfdualZ6}
\end{equation}
as a consequence of the chirality
condition of the spinors. We do not consider the commutation
relations between the bosonic generators. They are non-zero for
\OS, but zero for its contraction which, in $d=10$ and $d=11$,
is the super-\Poin\ algebra.
These additional commutation
relations do not play a role in the subsequent discussions.

In 11 dimensions, the reduction of the adjoint representation of
$Sp(32)$ goes as
\begin{equation}
(32\times 32)_S=11+55+462\,.
\label{number11}
\end{equation}
The relevant spinors are in an irreducible spinor representation of $SO(11)$.
The anticommutator is
\begin{equation}
\left\{Q,Q \right\} =\tilde{\Gamma }^{\tilde{M}}\tilde{Z}_{\tilde{M}}+
\ft12 \tilde{\Gamma
}^{\tilde{M}\tilde{N}}\tilde{Z}_{\tilde{M}\tilde{N}} +\ft1{5!}  \tilde{\Gamma
}^{\tilde{M}_1\cdots \tilde{M}_5}\tilde{Z}_{\tilde{M}_1\cdots
\tilde{M}_5}\,,
\label{calg11}
\end{equation}
where the gamma matrices are $32\times 32$. Their relation with the
12-dimensional ones is given in appendix~\ref{app:gamma}.

In 10 dimensions the smallest irreducible spinors are the 16-dimensional chiral
spinors. The 32 fermionic generators can be contained in 2 chiral spinors of
opposite chirality (IIA) or of the same chirality (IIB).
The 528 bosonic generators will be represented respectively as
\begin{eqnarray}
{\rm IIA} & : & 1+10+10+45+210+126^++126^-\, , \nonumber\\
{\rm IIB} & : & 10+10+10+120+126^++126^++126^+\,.
\label{number10}
\end{eqnarray}
For the IIA theory, the spinors satisfy $Q^\pm =\pm \Gamma_*Q^\pm$.
The IIA anticommutators are
\begin{eqnarray}
\left\{Q^{\pm },Q^{\pm } \right\} & = & {\cal P}^{\pm }\Gamma ^M Z^{\pm }_M
+\ft 1{5!}{\cal P}^{\pm }\Gamma ^{M_1\cdots M_5}Z^\pm _{M_1\cdots M_5}\, ,
\nonumber\\
\left\{Q^{\pm },Q^{\mp } \right\} & = & \pm {\cal P}^{\pm } Z +
\ft12{\cal P}^{\pm }\Gamma ^{MN}Z_{MN}
\pm \ft 1{4!}{\cal P}^{\pm }\Gamma ^{M_1\cdots M_4}Z _{M_1\cdots M_4}\,.
\label{calgIIA}
\end{eqnarray}
In this case $Z^\pm  _{M_1\cdots M_5}$ is (anti-)selfdual. For later
purposes, we note that
this algebra has the $ \Zbar _2$ automorphism which is denoted as $(-)^{\rm
F_L}$,
acting as
\begin{eqnarray}
Q^{\pm } & \stackrel{(-)^{\rm F_L}}\longrightarrow & \pm Q^\pm\, ,  \nonumber\\
\left( Z,Z_{MN},Z_{MNPQ}\right)  & \stackrel{(-)^{\rm F_L}}\longrightarrow  &
 -\left( Z,Z_{MN},Z_{MNPQ}\right)\,,
\label{mFL}
\end{eqnarray}
the other bosonic generators remaining invariant.

 For the IIB
theory we introduce a doublet of left-handed supersymmetries: $Q^i=\Gamma
_*Q^i$ with $i=1,2$, and the anticommutators are
\begin{equation}
  \left\{Q^i,Q^j \right\}={\cal P}^+\Gamma ^MY_M^{ij}
     +\ft1{3!}{\cal P}^+\Gamma ^{MNP}\varepsilon ^{ij}Y_{MNP}+
   {\cal P}^+\ft1{5!}\Gamma ^{M_1\cdots M_5 }Y_{M_1\cdots M_5 }^{+\,ij}
  \, ,
\label{calgIIB}
\end{equation}
where $Y_M^{ij}$ and $Y_{M_1\cdots M_5 }^{ij}$ are the most general
symmetric matrices in $(ij)$, which can be expanded in the $2\times 2$ Pauli
matrices as follows:
\begin{eqnarray}
Y_M^{ij} & = &  \delta ^{ij}Y_M^{(0)}
   +\tau _1^{ij}   Y_M^{(1)}+\tau _3^{ij}   Y_M^{(3)}\,, \nonumber\\
Y_{M_1\cdots M_5 }^{+\,ij} & = &  \left( \delta ^{ij}Y_{M_1\cdots M_5 }^{+(0)}
   +\tau _1^{ij}   Y_{M_1\cdots M_5 }^{+(1)}+\tau _3^{ij}
     Y_{M_1\cdots M_5 }^{+(3)}\right)\,.
\label{Y013}
\end{eqnarray}
This IIB algebra has $ \Zbar _2$
automorphisms, i.e.\ $Q^i\rightarrow M^i{}_j Q^j$, where $M$ is a
matrix that squares to $\unity $. In particular the cases $M=\tau _3$
and $M=\tau _1$ are usually denoted by $(-)^{\rm F_L}$ and $\Omega$,
which act as
\begin{eqnarray}
Q^i & \stackrel{(-)^{\rm F_L}}\longrightarrow  & (\tau  _3)^i{}_jQ^j\, , \nonumber\\
\left(Y^{(1)}_M, Y_{MNP}, Y^{+(1)}_{M_1\cdots M_5} \right)
 & \stackrel{(-)^{\rm F_L}}\longrightarrow
 & -\left(Y^{(1)}_M, Y_{MNP}, Y^{+(1)}_{M_1\cdots M_5} \right)\,,
\label{mFLII}\\
Q^i & \stackrel{\Omega }\longrightarrow  & (\tau  _1)^i{}_jQ^j\, ,  \nonumber\\
\left(Y^{(3)}_M, Y_{MNP}, Y^{+(3)}_{M_1\cdots M_5} \right)  & \stackrel{\Omega }\longrightarrow
 & -\left(Y^{(3)}_M, Y_{MNP}, Y^{+(3)}_{M_1\cdots M_5} \right)\,.
\label{FLIIOmega}
\end{eqnarray}
Finally, there is also the $S$ duality map
\begin{eqnarray}
Q^i & \stackrel{S}{\longrightarrow} & \left( e^{\rmi\ft14 \pi \tau _2}\right)
{}^i{}_jQ^j\, , \nonumber\\
\left( Y^{(1)}_M, Y^{(1)}_{M_1\cdots M_5}\right)
& \stackrel{S}{\longrightarrow} &
\left( Y^{(3)}_M, Y^{(3)}_{M_1\cdots M_5}\right)\, ,\nonumber\\
\left( Y^{(3)}_M, Y^{(3)}_{M_1\cdots M_5} \right)&
\stackrel{S}{\longrightarrow} &
-\left( Y^{(1)}_M, Y^{(1)}_{M_1\cdots M_5}\right)\,,
\label{S}
\end{eqnarray}
which has $S^8= 1 $.

\subsection{Dimensional reduction}
The algebras of the previous subsection can be related by dimensional
reduction, except for the IIB algebra which is related to the IIA by
T-duality, to be discussed in the next subsection.
The explicit formulae that we will give below, depend on the
representation of the Clifford algebra. We refer to appendix~\ref{app:gamma} for
our choice of representation.

The $d=12$ chiral spinor, $\hat{Q}$, is in this representation decomposed as
\begin{equation}
  \hat{Q}=\pmatrix{Q^+ \cr Q^- }\,.
\label{chiralitysplit}
\end{equation}
The chirality in 12 dimensions, with
$ \hat{\Gamma }_*=\Gamma _*\otimes \sigma _3$, implies that
the two components satisfy $Q^\pm =\pm \Gamma_*Q^\pm$. These
components are the supersymmetry generators of the
IIA algebra in 10 dimensions.
The 11-dimensional spinor $Q$ is obtained as the sum of $Q^+$ and $Q^-$:
\begin{equation}
Q=Q^++Q^-\,.
\label{Q11}
\end{equation}
It is easiest to reduce first from 12 to 10 dimensions. Then the IIA
algebra is obtained with
\begin{eqnarray}
Z_M^\pm  & = & \rmi\hat{Z}_{M\,12}\mp\hat{Z}_{M\,11}\, ,  \nonumber\\
Z_{M_1\cdots M_5} & = & 2 \rmi \hat{Z}_{M_1\cdots M_5\,12}\, ,
\nonumber\\
Z&=&\rmi \hat{Z}_{11\,12}\, ,\nonumber\\
Z_{MN}&=&\hat{Z}_{MN}\, ,\nonumber\\
Z_{M_1\cdots M_4}&=&2\rmi \hat{Z}_{M_1\cdots M_4\,11\,12}\, .
\label{ident1210}
\end{eqnarray}
After combining the anticommutators of $Q^\pm$ to obtain the anticommutation
relations for $Q$ as defined in (\ref{Q11}), we obtain
the following reduction rules from 12 to 11 dimensions:
\begin{eqnarray}
\tilde{Z}_{\tilde{M}} & = & \rmi \hat{Z}_{\tilde{M}\,12}\, , \nonumber\\
\tilde{Z}_{\tilde{M}\tilde{N}} & = & \hat{Z}_{\tilde{M}\tilde{N}}\, ,\nonumber\\
\tilde{Z}_{\tilde{M}_1\cdots \tilde{M}_5}&=&2\rmi \hat{Z}_{\tilde{M}_1\cdots
\tilde{M}_5\,12}\, .
\label{ident1211}
\end{eqnarray}
\subsection{T-duality}
\label{ss:T}
The connection between the IIA and IIB algebras is obtained by
T-duality in a particular direction. Since we work with the complex form of the
algebras as if we are in the Euclidean case, we denote this direction
by $s$ (spacelike). We make the following identifications between
the supersymmetry generators:
\begin{equation}
  Q^+= Q^1\,,\qquad Q^-=\Gamma^sQ^2\,.
\label{QIIAB}
\end{equation}
Splitting the 10-dimensional index $M$ in $(\mu ,s)$, with
$\mu = 1,\cdots ,9$, this leads to the following identifications
between the bosonic generators:
\begin{eqnarray}
Z_\mu ^\pm  & = & Y^{(0)}_\mu \pm  Y^{(3)}_\mu\, ,  \nonumber\\
Z_{s}^\pm   & = & \pm Y^{(0)}_{s} +  Y^{(3)}_{s}\, , \nonumber\\
Z_{\mu _1\cdots \mu _5}^\pm  & = & Y^{+(0)}_{\mu _1\cdots \mu _5}\pm
  Y^{+(3)}_{\mu _1\cdots \mu _5}\, , \nonumber\\
Z&=&-Y^{(1)}_{s}\, ,\nonumber\\
Z_{\mu \nu }&=&-Y_{\mu \nu \,s}\, ,\nonumber\\
Z_{\mu \,s}&=&-Y_\mu ^{(1)}\, ,\nonumber\\
Z_{\mu _1\cdots \mu _4}&=&-2Y^{+(1)}_{\mu _1\cdots \mu _4\,s}\, ,
\nonumber\\
Z_{\mu\nu \rho \,s}&=&-Y_{\mu\nu \rho }\, .
\label{Tdual}
\end{eqnarray}

For later purposes, to compare with the real case discussed in
Subsection~\ref{ss:Tdualityreal}, we explain how to derive the first two lines of
(\ref{Tdual}). We start by writing out the relevant terms of the IIA
algebra and split the index $M$ into $(\mu,s)$:
\begin{equation}
\{Q^\pm, Q^\pm\} = {\cal P}^\pm \Gamma^\mu
Z_\mu^\pm + {\cal P}^\pm \Gamma^s Z_s^\pm+\ldots \, .
\end{equation}
We next substitute the identifications (\ref{QIIAB}) into the left-hand-side
of this equation. We thus obtain:
\begin{eqnarray}
\{Q^1, Q^1\} &=& {\cal P}^+ \Gamma^\mu
Z_\mu^+ + {\cal P}^+ \Gamma^s Z_s^++\ldots\, ,\nonumber\\
\Gamma^s\{Q^2, Q^2\}(\Gamma^s)^T &=& {\cal P}^- \Gamma^\mu
Z_\mu^- + {\cal P}^- \Gamma^s Z_s^-+\ldots\, .
\end{eqnarray}
In the second line we bring the $\Gamma^s$ matrices to the right-hand-side
by using the identity $(\Gamma^s)^2 = 1$. Using some simple
gamma matrix identities like $C^{-1}(\Gamma^s)^T = - \Gamma^s C^{-1}$
(note again the omitted $C^{-1}$ in our notation for the anticommutation relations)
and $\Gamma^s \Gamma^\mu \Gamma^s = - \Gamma^\mu$, one finds
that the above algebra coincides with the relevant terms
in the IIB algebra provided the identifications are made given in
the first two lines of (\ref{Tdual}).

\section{Real symmetry algebras}\label{ss:realalgebras}
Up to now, we did not consider the hermiticity properties of the
generators. For the real forms of the algebras, we can impose
hermiticity conditions on the realizations of the generators. For the
fermionic generators, hermiticity conditions are the Majorana
conditions, whose consistency depends on the signature of spacetime
as mentioned in the introduction and discussed
in Appendix~\ref{app:conventions}. The situation
is summarized in Table~\ref{tbl:MWSspinors}.
\begin{table}[ht]
\begin{center}
\begin{tabular}{|c|cccccc|}
\hline
$d$  &  \multicolumn{6}{c|}{$(s,t)$}    \\
\hline
12  & (10,2) &  &  &  & (6,6) &  \\
    & MW     &  &  &  & MW &  \\
\hline
11  & (10,1)  & (9,2) &  &  & (6,5) &  \\
    & M       & M     &  &  &  M &  \\
\hline
10  & (10,0) & (9,1)  & (8,2) & (7,3) & (6,4) & (5,5) \\
    & SM      & MW     & M     & SMW   &  SM     & MW \\
\hline
    & A      & A/B     &   A   &  B    &  A    & A/B\\
\hline
\end{tabular}
  \caption{\sl The Table summarizes the possible reality
conditions for spinors in dimensions $d=10,11,12$ for different
  signatures $(s,t)$ in which the minimal spinor is at most 32-dimensional.
 M stands for Majorana spinors,  SM is a shorthand for
symplectic Majorana spinors (given our convention for charge conjugation,
otherwise this is equivalent to M), MW indicates the
possibility of Majorana--Weyl spinors, while SMW indicates the possibility of
symplectic Majorana--Weyl spinors. We only give the signatures with
$s\geq t$, as those with $s<t$ are equivalent up to interchange of
$s$ with $t$. The last row indicates, for each signature,
whether in $d=10$ a real form for Type
IIA (A), Type IIB (B) or both (A/B) exists.
 }\label{tbl:MWSspinors}
\end{center}
\end{table}

\subsection{Algebras in dimensions 12 to 10}
To consider the hermiticity properties of the generators, it is
convenient to replace complex conjugation by the operation of charge
conjugation\footnote{As can be seen from appendix~\ref{app:conventions},
the charge conjugation operation on spinors squares to one in the signatures
that appear in Table~\ref{tbl:MWSspinors}, except for $(10,0)$,
$(7,3)$ and $(6,4)$.
For $(10,0)$ and $(6,4)$ one could define another charge conjugation
matrix, multiplying the one we defined in appendix~\ref{app:conventions}
by $\Gamma _*$, such that the operation
with respect to the newly defined charge conjugation
matrix also squares to one. The charge conjugation matrix
which we use in this paper leads in the latter signatures
to a symplectic Majorana condition as discussed in the text.\label{fn:C}}.
This operator is chosen such that Majorana spinors
$\lambda $ satisfy $ \lambda ^C=\lambda $. For more details, see
Appendix~\ref{app:conventions}. The definition of the $C$
operation involves an arbitrary phase factor $\alpha $ (or a matrix,
see (\ref{lambdaC}))
which may depend on the spacetime dimension and signature.
This phase factor is equivalent to a possible redefinition of all
spinors by $\alpha ^{1/2}$.
For bosonic generators, the
$C$ operation is hermitian conjugation together with
multiplication with a factor of $\alpha^{-2} \beta$ (see
(\ref{B*})).
The parameter $\beta$ is a sign that depends on the convention whether one
maintains ($\beta=1$) or interchanges ($\beta = -1$) the order of fermions.
Finally, for $\Gamma $-matrices
the important properties are
\begin{equation}
  \Gamma _a ^C= (-)^{t+1}\Gamma _a\,,\qquad \Gamma _*^C=(-)^{d/2+t}\Gamma
  _*\,,
\label{GammaC}
\end{equation}
where $a$ denotes a Lorentz index in an arbitrary dimension.

To connect gamma matrices in different spacetime signatures, we use
for gamma matrices in timelike directions
\begin{equation}
  \Gamma ^a=-\rmi \Gamnul{}^a\,,
\label{Gammatimelike}
\end{equation}
where $ \Gamnul$ denotes the gamma matrix in Euclidean
space, which was used in the complex form of the algebras.
We will not change the form of the commutators of the previous
section, by redefining simultaneously the generators, e.g.
\begin{equation}
   \Gamnul{} ^{13}\Znul  _{13}={\Gamma}
   ^{13}{Z}_{13}\qquad \mbox{with}\qquad{Z}_{13}=\rmi
   \Znul_{13}\,,
\label{redefZ}
\end{equation}
for 1 being a timelike direction, and 3 a spacelike direction.

Considering the consistency of the anticommutators of the previous
section with the charge conjugation, we find for $d=12$
\begin{equation}
  d=12\ :  \hat{Z}= \hat{Z}^\ncc
\label{herm12}
\end{equation}
for all generators. We have introduced here a $\ncc$-operation for the
bosonic generators that basically is a complex conjugation that
hides prefactors involving $\alpha$ and $\beta$
(see eq.~\ref{B*}).
For $d=12$ the $\ncc$-operation involves a factor $\alpha_{10,2} ^{-2}\beta$
or $\alpha_{6,6} ^{-2}\beta$. The factor $\alpha
$ is in this case a number with modulus one, that can be chosen for
each spacetime signature.

Similarly, for $d=11$ and $t=1$ we find the reality conditions:
\begin{equation}
  d=11\,,\ t=1,5\ :\qquad\tilde Z=\tilde{Z}^\ncc\,,
\label{herm11t1}
\end{equation}
for all bosonic generators $\tilde Z$ with 1, 2 or 5 indices.
For two time directions in $d=11$, other signs occur:
\begin{equation}
   d=11\,,\ t=2\ :\qquad
\begin{array}{l}
  \tilde Z_{\tilde M}=-\tilde{Z}^\ncc_{\tilde M}\,, \\
 \tilde Z_{\tilde M\tilde N}=\tilde{Z}^\ncc_{\tilde M\tilde N}\,, \\
  \tilde Z_{\tilde M_1\cdots \tilde M_5}=-\tilde{Z}^\ncc_{\tilde M_1\cdots
\tilde M_5}\,.\end{array}
\label{herm11t2}
\end{equation}

For 10 dimensions the discussion becomes more involved as there are
different types of spinors. Consider first the signatures $(9,1)$ and
$(5,5)$, where they are Majorana--Weyl spinors
\begin{eqnarray}
&&{\rm IIA}\,,\ (9,1)\mbox{ or }(5,5)\ :\qquad
(Q^+)^C=Q^+\,,\qquad (Q^-)^C=Q^-\, , \nonumber\\
&&{\rm IIB}\,,\ (9,1)\mbox{ or }(5,5)\ :\qquad   (Q^i)^C=Q^i\,.
\label{QMaj91}
\end{eqnarray}
This is possible because $\Gamma _*$ is invariant under $C$. We could
have inserted arbitrary phase factors in the right-hand sides of
these equations. However, these amount to a redefinition of the phase
factor $\alpha $ in the definition of the charge conjugation of
spinors. This is true for any Majorana condition. For IIA
we have two independent Majorana conditions,
and correspondingly we could have two independent phase factors hidden
in the $C$ operation. In other words, the matrix $\alpha $ is in this
case
\begin{equation}
\alpha^{(A)}_{9,1} = \pmatrix{\alpha_{9,1}^+& 0\cr
                               0& \alpha_{9,1}^-}\, .
\label{alphaA91}
\end{equation}
For IIB the Majorana condition can mix the two chiral spinors, and
$\alpha$ can thus be a matrix $\alpha ^i{}_j$, that has to satisfy
(\ref{alphaalpha*}). For clarity, we
give below the explicit form of the IIB reality condition given in
the second line of (\ref{QMaj91}):
\begin{equation}
(Q^i)^C \equiv (\alpha^{-1})^i{}_j\,  \Gamma_t C^{-1} Q^{j*} = Q^i\, .
\label{Qstar}
\end{equation}

One obtains for all bosonic
generators in the superalgebra the reality conditions
\begin{eqnarray}
  &   & {\rm IIA}\,,\ (9,1)\mbox{ or }(5,5)\ :\qquad  Z= Z^\ncc \,, \nonumber\\
&&{\rm IIB}\,,\ (9,1)\mbox{ or }(5,5)\ :\qquad Y= Y^\ncc\, .
\label{hermZ91}
\end{eqnarray}
In the first equation it is understood that, in the case that one
uses different phase factors $\alpha^\pm$ for the reality conditions
on $Q^\pm$ the $\ncc$-operation reads:
\begin{eqnarray}
\left (Z^\pm\right )^\ncc &=&  \beta\, (\alpha^\pm )^{-2} (Z^\pm)^*\, ,
\nonumber\\
Z^\ncc &=& \beta\,(\alpha^+\alpha^-)^{-1} Z^*\, ,
\label{r}
\end{eqnarray}
where the second line is for the Ramond--Ramond operators.
In the second equation of (\ref{hermZ91}), the implicit $\alpha $ factors act now as a
matrix, thus
\begin{equation}
Y^{ij}=  Y^{\ncc\,ij}= \beta (\alpha^{-1}) ^i{}_k\, Y^{* \, k\ell }\,
  (\alpha^{-1}) ^j{}_\ell \,,\qquad
\varepsilon^{ij}Y_{MNP}= \beta (\alpha^{-1}) ^i{}_k\,\varepsilon ^{k\ell}Y_{MNP}^{* }\,
  (\alpha^{-1}) ^j{}_\ell     \,.
\label{Ystar}
\end{equation}

Next, we consider the signature (8,2).
The Majorana condition can only be imposed
on non-chiral spinors. Indeed, the $C$ operation changes a chiral
spinor to an antichiral one as $\Gamma _*^C=-\Gamma _*$. Therefore,
we can not impose Majorana conditions on chiral spinors, and thus
type IIB does not have a real form. For type IIA, the Majorana spinor
is $Q^++Q^-$, or explicitly we have
\begin{equation}
{\rm  IIA}\,,\ (8,2)\ :\qquad    (Q^+)^C=Q^-\,,\qquad (Q^-)^C=Q^+\,.
\label{Q+-C}
\end{equation}
Considering here the consistency of the algebra with the $C$
operation, we find
\begin{eqnarray}
{\rm  IIA}\,,\ (8,2)\ :\qquad  &&(Z_M^+)^\ncc  =  -Z_M^-\,,\qquad
(Z_{M_1\cdots M_5 }^+)^\ncc  =  -Z_{M_1\cdots M_5 }^-\, , \nonumber\\
&&Z^\ncc  =  -Z\,,\qquad Z_{MN}^\ncc  =  Z_{MN}\,,\qquad Z_{MNPQ}^\ncc  =
-Z_{MNPQ}\,.
\label{hermZ82}
\end{eqnarray}

For the signatures $(10,0)$ and $(6,4)$ the discussion is similar to
the $(8,2)$ case. Again, a real form of type IIB does not exist. For type
IIA there are only signs to be changed with respect to the
previous case. Indeed, as mentioned in
footnote~\ref{fn:C} by another definition of charge conjugation the
(10,0) and (6,4) cases could be made identical to the (8,2) case.
However, we do not change our definition of the
charge conjugation, which implies that now the charge conjugation
operator squares to $-1$. Therefore, a consistent Majorana condition
looks rather as a symplectic Majorana condition:
\begin{equation}
{\rm  IIA}\,,\ (10,0)\mbox{ or }(6,4)\ :\qquad   (Q^+)^C=Q^-\,,\qquad (Q^-)^C=-Q^+\,.
\label{Q+-CE}
\end{equation}
The consistency of the algebra with the $C$
operation gives now other signs for the Ramond--Ramond generators
\begin{eqnarray}
{\rm  IIA}\,,\ (10,0)\mbox{ or }(6,4)\ :\qquad &&(Z_M^+)^\ncc  =  -Z_M^-\,,
 \qquad (Z_{M_1\cdots M_5 }^+)^\ncc  =  -Z_{M_1\cdots M_5 }^-\, ,
\label{hermZE} \\
&&Z^\ncc  =  Z\,,\qquad Z_{MN}^\ncc  = - Z_{MN}\,,\qquad Z_{MNPQ}^\ncc  =
Z_{MNPQ}\,.\nonumber
\end{eqnarray}

The remaining case is $(7,3)$. Now, the charge conjugation leaves $\Gamma
_*$ invariant, and squares to $-1$. Therefore in type IIA we can not
impose Majorana conditions, or, in other words, there is no real form
for IIA. For type IIB there is the symplectic Majorana--Weyl condition
\begin{equation}
 IIB\,,\ (7,3)\ :\qquad   (Q^i)^C =\varepsilon ^i{}_j Q^j\,.
\label{symplMajIIB}
\end{equation}
The reality conditions of the bosonic generators are
\begin{eqnarray}
{\rm  IIB}\,,\ (7,3)\ :\qquad   &(Y_M^{(0)})^\ncc  = Y_M^{(0)}\,,&\qquad
 (Y_{M_1\cdots M_5 }^{+(0)})^\ncc  = Y_{M_1\cdots M_5 }^{+(0)}
 \,, \nonumber\\
 & (Y_M^{(1)})^\ncc  = -Y_M^{(1)}\,,& \qquad
 (Y_{M_1\cdots M_5 }^{+(1)})^\ncc  = -Y_{M_1\cdots M_5 }^{+(1)}\,, \nonumber\\
& (Y_M^{(3)})^\ncc  = -Y_M^{(3)}\,,&\qquad
 (Y_{M_1\cdots M_5 }^{+(3)})^\ncc  = -Y_{M_1\cdots M_5 }^{+(3)}\, ,\nonumber\\
&(Y_{MNP})^\ncc = Y_{MNP}\,.&
\label{realityY73}
\end{eqnarray}
\vspace{3mm}

Finally, we consider the automorphisms mentioned in
subsection~\ref{ss:calg1210}.
For type IIA, the automorphism $(-)^{\rm F_L}$, (\ref{mFL}), is only preserved for the
signatures (9,1) and (5,5). In these cases a projection to the even
part under this automorphism leads to the $N=1$ algebra, i.e.\
$OSp(1|16)$. Also for type IIB, the automorphisms $(-)^{\rm F_L}$
 and $\Omega $, see (\ref{mFLII}) and
(\ref{FLIIOmega}), are only preserved for
these signatures. The type IIA and type IIB
automorphisms for the (9,1) signature have
been discussed in \cite{Townsend:1997wg}.
The $S$ duality is preserved in all real forms of
IIB, i.e.\ also for (7,3).

\subsection{Dimensional reduction}

We start with the 12-dimensional Majorana condition for (10,2)
signature,
\begin{equation}
  \hat{Q}=\frac{1}{\alpha _{10,2}}\hat{B}^{-1}\hat{Q}^*\,,
  \qquad\hat{B}^{-1}=\hat{\Gamma }_{t_1}\hat{\Gamma }_{t_2}\hat{C}{}^{-1}\,,
\label{hatQMaj}
\end{equation}
where we have indicated the signature for the phase factor explicitly.
We now
reduce this to 10 dimensions. There are different ways to do so,
related to the identification of the 12-dimensional timelike directions
$t_1$ and $t_2$ with either timelike directions in the 10 uncompactified
dimensions or
the two compactified dimensions. Neglecting the order of the two (which is
another sign factor\footnote{A sign change of $\alpha $ amounts
to a redefinition of the fermionic generators with $\rmi$ and of the
bosonic generators with a sign.} in $\alpha $), there are 4 different
choices. If the compactification directions are 11 and 12, and we use
the representation of 12-dimensional gamma matrices as in
(\ref{Gamma64}), inserting factors $\rmi$ for timelike directions,
we obtain 10-dimensional Majorana conditions. Comparing these
with those in the previous subsection, we find the following
relations between the phase factors in ten and twelve dimensions:
\begin{eqnarray}
t_1=11,\ t_2=12 & (\mbox{via }(10,1))\ : & \alpha _{10,0}= -\rmi \alpha _{10,2}\, , \nonumber\\
t_1=\phantom{1}1,\ t_2=11 & (\mbox{via }(\phantom{1}9,2))\ : & \alpha _{9,1}^\pm =\pm\rmi \alpha _{10,2}\, , \nonumber\\
t_1=\phantom{1}1,\ t_2=12 & (\mbox{via }(10,1))\ : & \alpha _{9,1}^\pm=-\alpha _{10,2}\, , \nonumber\\
t_1=\phantom{1}1,\ t_2=\phantom{1}2 &(\mbox{via }(\phantom{1}9,2))\  : & \alpha _{8,2}= \alpha _{10,2}\,.
\label{resultalpha10}
\end{eqnarray}
Observe that the results for signature (9,1), which are obviously type IIA
theories, depend on whether the
first compactification direction (direction 12) is spacelike, and the
second one (direction 11) timelike, as it is in the second line, or
the reverse as in the third line.

The above rules apply as well to the reduction of the $(6,6)$ spinors.
In that case
the expressions for $\alpha $ are obtained from the above ones by
replacing $\alpha _{s,t}$ by $\alpha _{s-4,t+4}$, e.g.\ the first line
in (\ref{resultalpha10}) gives $\alpha_{6,4}=-\rmi \alpha _{6,6}$.

We can combine the 10-dimensional Majorana spinors to 11-dimensional
ones. The 11-dimensional spinor is given by (\ref{Q11}). In the above
scheme, the (10,1) theory where `1' is the time direction can be
obtained from the third line in (\ref{resultalpha10}). One can check
that this leads to the Majorana condition of (10,1) with
\begin{equation}
  \alpha _{10,1}= \alpha ^\pm _{9,1}=-\alpha _{10,2}\,.
\label{alpha101}
\end{equation}
In principle, one could also start from the first line in
(\ref{resultalpha10}).

For the (9,2) theory, we have standard the timelike directions as `1'
and `2'. Thus this can be obtained from the fourth line of
(\ref{resultalpha10}). Combining the corresponding chiral spinors
leads to a (9,2) Majorana spinor with
\begin{equation}
  \alpha _{9,2}=\alpha _{8,2}=\alpha _{10,2}\,.
\label{alpha92}
\end{equation}

To obtain the dimensional reduction of the bosonic generators, we
start from those in the complex case, but we have to take into
account extra factors of $\rmi$ due to (\ref{redefZ}). For instance,
when we reduce
from (10,2) to (10,1), we start from (\ref{ident1211}), which is a formula for
the $\Znul$, i.e.~this formula refers to the Euclidean case.
Then taking into account that the twelfth direction is a
time direction, the $\rmi$ factors cancel for the relation between
the $Z$ generators:
\begin{eqnarray}
 \tilde{Z}_{\tilde{M}} & = & \hat{Z}_{\tilde{M}\,12}\, , \nonumber\\
(10,2)\rightarrow (10,1)\ :\qquad\tilde{Z}_{\tilde{M}\tilde{N}} & = &
\hat{Z}_{\tilde{M}\tilde{N}}\, ,\nonumber\\
\tilde{Z}_{\tilde{M}_1\cdots \tilde{M}_5}&=&2 \hat{Z}_{\tilde{M}_1\cdots
\tilde{M}_5\,12}\,.
\label{ident102101}
\end{eqnarray}
On the other hand, when we reduce from (10,2) to (9,2) the twelfth direction is
spacelike, so there are no extra factors of $\rmi$ when we replace $\Znul$
by $Z$ in (\ref{ident1211}):
\begin{eqnarray}
\tilde{Z}_{\tilde{M}} & = & \rmi \hat{Z}_{\tilde{M}\,12}\, , \nonumber\\
(10,2)\rightarrow (9,2)\ :\qquad\tilde{Z}_{\tilde{M}\tilde{N}} & = &
\hat{Z}_{\tilde{M}\tilde{N}}\, ,\nonumber\\
\tilde{Z}_{\tilde{M}_1\cdots \tilde{M}_5}&=&2\rmi \hat{Z}_{\tilde{M}_1\cdots
\tilde{M}_5\,12}\,.
\label{ident10292}
\end{eqnarray}
One can check the consistency of these rules with the reality
properties of the previous subsection,
see (\ref{herm12}), (\ref{herm11t1}) and (\ref{herm11t2}).
One has to keep in mind that the $\ncc$-operation involves factors of $\alpha
$, see (\ref{B*}). In this case
$\alpha_{10,2} ^2= \alpha ^2_{10,1}=\alpha ^2_{9,2}$, so that this
does not lead to a change of sign in the reality property.

The reduction from 12 to 10 can be done in 4 different ways. With
time directions 11 and 12 we get
\begin{eqnarray}
Z_M^\pm  & = & \hat{Z}_{M\,12}\pm\rmi\hat{Z}_{M\,11}\, ,  \nonumber\\
Z_{M_1\cdots M_5} & = & 2  \hat{Z}_{M_1\cdots M_5\,12}\, ,
\nonumber\\
(10,2)\rightarrow (10,0)\ :\qquad\qquad Z&=&-\rmi \hat{Z}_{11\,12}\, ,\nonumber\\
Z_{MN}&=&\hat{Z}_{MN}\, ,\nonumber\\
Z_{M_1\cdots M_4}&=&-2\rmi \hat{Z}_{M_1\cdots M_4\,11\,12}\,.
\label{ident12t1112}
\end{eqnarray}
With time directions 1 and 11 we get
\begin{eqnarray}
Z_M^\pm  & = & \rmi(\hat{Z}_{M\,12}\pm\hat{Z}_{M\,11})\, ,  \nonumber\\
Z_{M_1\cdots M_5} & = & 2 \rmi \hat{Z}_{M_1\cdots M_5\,12}\, ,
\nonumber\\
(10,2)\rightarrow (9,1)\ :\qquad\qquad Z&=& \hat{Z}_{11\,12}\, ,\nonumber\\
Z_{MN}&=&\hat{Z}_{MN}\, ,\nonumber\\
Z_{M_1\cdots M_4}&=&2 \hat{Z}_{M_1\cdots M_4\,11\,12}\,.
\label{ident12t111}
\end{eqnarray}
Note that to check the reality properties in this case, one has to
take into account that for the Ramond--Ramond generators, the
$\ncc$-operation involves $\alpha ^+_{9,1}\alpha^-_{9,1}$, see (\ref{r}).
For time directions 1 and 12 we get
\begin{eqnarray}
Z_M^\pm  & = & \hat{Z}_{M\,12}\mp\hat{Z}_{M\,11}\, ,  \nonumber\\
Z_{M_1\cdots M_5} & = & 2  \hat{Z}_{M_1\cdots M_5\,12}\, ,
\nonumber\\
(10,2)\rightarrow (9,1)\ :\qquad\qquad Z&=& \hat{Z}_{11\,12}\, ,\nonumber\\
Z_{MN}&=&\hat{Z}_{MN}\, ,\nonumber\\
Z_{M_1\cdots M_4}&=&2 \hat{Z}_{M_1\cdots M_4\,11\,12}\,.
\label{ident12t112}
\end{eqnarray}
The two (9,1) algebras we obtain from (\ref{ident12t111}) and
(\ref{ident12t112}) are related to each other via a simple redefinition.
To exhibit this redefinition, it is convenient
to denote the (9,1) generators obtained from (\ref{ident12t111}) with
Q and Z while those obtained from (\ref{ident12t112}) with $\tilde Q$ and
$\tilde Z$. Using this notation we find
\begin{eqnarray}
Q^+ &=& \rmi^{1/2} {\tilde Q}^-\, ,\hskip 1.5truecm
Q^- = (-\rmi)^{1/2} {\tilde Q}^+\, ,
\nonumber\\
-\rmi Z_M^+ &=& {\tilde Z}_M^-\, ,\hskip 2truecm
\rmi Z_M^- = {\tilde Z}_M^+\, ,\nonumber\\
-\rmi Z_{M_1\cdots M_5}^+ &=& {\tilde Z}_{M_1\cdots M_5}^-\, ,
\hskip .5truecm
\rmi Z_{M_1\cdots M_5}^- = {\tilde Z}_{M_1\cdots M_5}^+\, ,\\
Z &=& - {\tilde Z}\, ,\hskip 2truecm Z_{MN} = {\tilde Z}_{MN}\, ,\nonumber\\
Z_{M_1\cdots M_4} &=& - {\tilde Z}_{M_1\cdots M_4}\, .\nonumber
\end{eqnarray}
This redefinition is consistent with the twelve-dimensional
origin of the generators given in (\ref{ident12t111}) and
(\ref{ident12t112}) provided we interchange in the latter equation
the 11 and 12 directions.

Finally, for time directions 1 and 2 there is no change with respect
to the rules of the complex algebra:
\begin{eqnarray}
Z_M^\pm  & = & \rmi\hat{Z}_{M\,12}\mp\hat{Z}_{M\,11}\, ,  \nonumber\\
Z_{M_1\cdots M_5} & = & 2 \rmi \hat{Z}_{M_1\cdots M_5\,12}\, ,
\nonumber\\
(10,2)\rightarrow (8,2)\ :\qquad\qquad Z&=&\rmi \hat{Z}_{11\,12}\, ,\nonumber\\
Z_{MN}&=&\hat{Z}_{MN}\, ,\nonumber\\
Z_{M_1\cdots M_4}&=&2\rmi \hat{Z}_{M_1\cdots M_4\,11\,12}\,.
\label{ident12t12}
\end{eqnarray}

\subsection{T-duality}
\label{ss:Tdualityreal}

We first discuss the conventional \textit{space/space} dualities. These dualities
do not change the signature of spacetime and only exist for the
signatures (9,1) and (5,5) where both a IIA and a IIB algebra can be defined
(see also Table~\ref{tbl:MWSspinors}):
\begin{equation}
{\rm IIA}_{(9,1)} \stackrel{{\rm T_{ss}}}{\longleftrightarrow} {\rm IIB}_{(9,1)}
\, ,\hskip 1truecm
{\rm IIA}_{(5,5)} \stackrel{\rm T_{ss}}{\longleftrightarrow} {\rm IIB}_{(5,5)}\, .
\end{equation}
 The identification between the supersymmetry generators
is as in the complex case, see \eqs (\ref{QIIAB}) and one obtains the same
T-duality rules as in the complex case, see \eqs (\ref{Tdual}). Let us
check that these T-duality rules are consistent with the reality
properties of the different generators. Starting
with the reality conditions on the IIA supersymmetry generators as given in
the first line of \eqs (\ref{QMaj91}) and applying the T-duality rule
(\ref{QIIAB}) one finds the reality condition (\ref{Qstar}) on the
IIB supergenerators with the matrix $\alpha^{(B)}_{9,1}$
\begin{equation}
  \alpha ^{(B)}_{9,1}=\alpha^{(A)}_{9,1} =
  \pmatrix{\alpha_{9,1}^+& 0\cr
                               0& \alpha_{9,1}^-}\, .
\label{alphaB91}
\end{equation}
Applying (\ref{Ystar}), this leads to the following reality
conditions on the bosonic generators of the IIB (9,1) algebra:
\begin{eqnarray}
\left(Y_M^{(0)} \pm Y_M^{(3)}\right )^* &=& \beta(\alpha^\pm_{9,1})^{2}
\left (Y_M^{(0)} \pm Y_M^{(3)}\right )\, ,\nonumber\\
\left (Y_M^{(1)}\right )^* &=& \beta \alpha^+_{9,1}\alpha_{9,1}^-
 Y_M^{(1)}\, ,\\
\left(Y_{M_1\cdots M_5}^{(0)} \pm Y_{M_1\cdots M_5}^{(3)}\right )^*&=&
\beta(\alpha^\pm_{9,1})^2
\left (Y_{M_1\cdots M_5}^{(0)} \pm Y_{M_1\cdots M_5}^{(3)}
\right )\, ,\nonumber\\
Y_{MNP}^* &=& \beta \alpha^+_{9,1}\alpha_{9,1}^- Y_{MNP}\, .
\label{realityAB}
\end{eqnarray}
One may  verify that the above reality conditions on the bosonic
IIB generators and the corresponding conditions (\ref{r}) on the bosonic
IIA generators are consistent with the T-duality rules (\ref{Tdual}).

Also \textit{time/time} dualities give a relation between the IIA and
IIB (9,1) algebras.
In this case the duality is performed in a
timelike direction $t$. To apply the rules of Subsection~\ref{ss:T},
 we write the gamma matrix
$\Gamma^s$ as $\Gamma^s = \rmi \Gamma^t$ such that $(\Gamma^s)^2 =
(\rmi\Gamma^t)^2 =1$. We thus make the following identifications between
the supersymmetry generators:
\begin{equation}
  Q^+= Q^1\,,\qquad Q^-= \rmi\Gamma^tQ^2\,,
\label{QIIABtt}
\end{equation}
The calculation for the spacelike case can thus be copied, replacing
everywhere $\Gamma^s$ by $\rmi \Gamma^t$ and $s$-like components of
the bosonic charges by $-\rmi$ times the $t$-like components as explained in
\eqs (\ref{redefZ}). We thus we find the following result,
\begin{eqnarray}
Z_\mu ^\pm  & = & Y^{(0)}_\mu \pm  Y^{(3)}_\mu\, ,  \nonumber\\
Z_{t}^\pm   & = & \left (\pm Y^{(0)}_{t} +  Y^{(3)}_{t}\right)
\, , \nonumber\\
Z_{\mu _1\cdots \mu _5}^\pm  & = & Y^{+(0)}_{\mu _1\cdots \mu _5}\pm
  Y^{+(3)}_{\mu _1\cdots \mu _5}\, , \nonumber\\
{\rm IIA}, (9,1) \stackrel{\textrm{T}_{tt}}{\longleftrightarrow} {\rm IIB}, (9,1)\ :\qquad\qquad
Z&=&\rmi Y^{(1)}_{t}\, ,\nonumber\\
Z_{\mu \nu }&=&\rmi Y_{\mu \nu \,t}\, ,\nonumber\\
Z_{\mu \,t}&=&\rmi Y_\mu ^{(1)}\, ,\nonumber\\
Z_{\mu _1\cdots \mu _4}&=&2\rmi Y^{+(1)}_{\mu _1\cdots \mu _4\,t}\, ,
\nonumber\\
Z_{\mu\nu \rho \,t}&=&\rmi Y_{\mu\nu \rho }\, .
\label{Tdualtt}
\end{eqnarray}

Again, these T-duality rules should be consistent with the reality
conditions of the generators discussed before. The reality condition
of the IIA supersymmetry generators is determined by (\ref{QMaj91}),
in terms of the matrix $\alpha ^{(A)}_{9,1}$. Calculating then
$Q^{i*}$, we find
\begin{equation}
  Q^{1*}=B\alpha ^+_{9,1}Q^1\,,\qquad Q^{2*}= -B\alpha ^-_{9,1}Q^2\,,
\label{Q12*tt}
\end{equation}
which are the IIB reality conditions of (\ref{QMaj91}) when we take
\begin{equation}
  \alpha ^{(B)}_{9,1}=\pmatrix{\alpha ^+_{9,1}&0\cr 0&-\alpha
  ^-_{9,1}}\,.
\label{alphatt}
\end{equation}
Given this $\alpha ^{(B)}$ one can check that the time/time duality
rules (\ref{Tdualtt}) are
 consistent with the reality conditions of the real IIA (9,1)
and IIB (9,1) algebras.

To explain how the \textit{space/time} T-duality between the real algebras works, we
first consider the identification between the IIA (10,0) and the
IIB (9,1) algebras. Note that we are dealing with a space/time T-duality,
i.e.~in the IIA (10,0) algebra the duality is performed in a
spacelike direction $s$ whereas in the IIB (9,1) algebra we perform
T-duality in the single timelike direction $t$. Again, we can copy
the results for Subsection~\ref{ss:T}, this time we only have to insert
$\rmi$ factors on the IIA side as explained for the time/time duality above.
We thus find
\begin{eqnarray}
Z_\mu ^\pm  & = & Y^{(0)}_\mu \pm  Y^{(3)}_\mu\, ,  \nonumber\\
Z_{s}^\pm   & = & -\rmi \left (\pm Y^{(0)}_{t} +  Y^{(3)}_{t}\right)
\, , \nonumber\\
Z_{\mu _1\cdots \mu _5}^\pm  & = & Y^{+(0)}_{\mu _1\cdots \mu _5}\pm
  Y^{+(3)}_{\mu _1\cdots \mu _5}\, , \nonumber\\
{\rm IIA}, (10,0) \stackrel{\textrm{T}_{st}}{\longleftrightarrow} {\rm IIB}, (9,1)\ :\qquad\qquad
Z&=&\rmi Y^{(1)}_{t}\, ,\nonumber\\
Z_{\mu \nu }&=&\rmi Y_{\mu \nu \,t}\, ,\nonumber\\
Z_{\mu \,s}&=&-Y_\mu ^{(1)}\, ,\nonumber\\
Z_{\mu _1\cdots \mu _4}&=&2\rmi Y^{+(1)}_{\mu _1\cdots \mu _4\,t}\, ,
\nonumber\\
Z_{\mu\nu \rho \,s}&=&-Y_{\mu\nu \rho }\, .
\label{Tdual10,0}
\end{eqnarray}
The reality condition
of the (10,0) supersymmetry generators is, according to \eqs (\ref{Q+-CE}),
given by
\begin{equation}
{-1\over \alpha_{10,0}}C^{-1} (Q^+)^* = Q^-\, ,\hskip 2truecm
{-1\over \alpha_{10,0}}C^{-1} (Q^-)^* = -\, Q^+\, ,\end{equation}
where $\alpha_{10,0}$ is an arbitrary phase factor. The reality conditions
of the (10,0) bosonic generators are given in (\ref{hermZE}).
Applying the T-duality relation (\ref{QIIABtt}) on the IIA
supersymmetry generators we find the
reality condition (\ref{Qstar}) on the IIB supersymmetry generators
with the matrix $\alpha^{(B)}_{9,1}$ given by
\begin{equation}
\alpha^{(B)}_{9,1} = -\rmi\, \alpha_{10,0}\, \tau_1\, .
\label{alpha91100}
\end{equation}
Using this expression we can write out \eqs (\ref{Ystar})
in components. This leads to the following reality
conditions on the bosonic generators of the IIB algebra:
\begin{eqnarray}
\label{reality9,1}
Y_M^{(0)\,*} &=& -\beta\alpha_{10,0}^2\,Y_M^{(0)}
\, , \qquad
Y_{M_1\cdots M_5}^{(0)\,*} =
-\beta\alpha_{10,0}^2\,Y_{M_1\cdots M_5}^{(0)}\, ,
\nonumber\\
Y_M^{(1)\,*} &=& -\beta \alpha_{10,0}^2\, Y_M^{(1)}\, ,
\qquad
Y_{M_1\cdots M_5}^{(1)\,*} = -\beta \alpha_{10,0}^2\, Y_{M_1\cdots M_5}^{(1)}\, ,
\nonumber\\
Y_M^{(3)\,*}&=&\phantom{-} \beta\alpha_{10,0}^2\, Y_M^{(3)}\,,\qquad
Y_{M_1\cdots M_5}^{(3)\,*} =\phantom{-}
\beta\alpha_{10,0}^2\, Y_{M_1\cdots M_5}^{(3)}
\, ,\nonumber\\
Y_{MNP}^* &=& \phantom{-}\beta \alpha_{10,0}^2\, Y_{MNP}\, .
\end{eqnarray}
It is now straightforward to verify that the reality conditions
(\ref{hermZE}) and (\ref{reality9,1}) are consistent with the
T-duality rules (\ref{Tdual10,0}) between the real IIA (10,0) and IIB
(9,1) superalgebras.

The T-duality between the real IIA (10,0) and IIB (9,1)
algebras can be extended to the following chain of
space/time T-dualities:
\begin{equation}
{\rm IIA}_{(10,0)} \stackrel{{\rm T}_{st}}{\longleftrightarrow}
{\rm IIB}_{(9,1)} \stackrel{{\rm T}_{st}}{\longleftrightarrow}
{\rm IIA}_{(8,2)} \stackrel{{\rm T}_{st}}{\longleftrightarrow}
{\rm IIB}_{(7,3)} \stackrel{{\rm T}_{st}}{\longleftrightarrow}
{\rm IIA}_{(6,4)} \stackrel{{\rm T}_{st}}{\longleftrightarrow}
{\rm IIB}_{(5,5)}\, .
\end{equation}
Consider for instance the IIA (8,2) supergenerators. They satisfy
the following reality conditions:
\begin{equation}
{1\over \alpha_{8,2}}\Gamma_1\Gamma_2 C^{-1} (Q^+)^* = Q^-\, ,\hskip 2truecm
{1\over \alpha_{8,2}}\Gamma_1\Gamma_2C^{-1} (Q^-)^* = \, Q^+\, ,
\end{equation}
where $\alpha_{8,2}$ is an arbitrary phase factor and 1 and 2 are the two
timelike directions. Performing a T-duality in one of the timelike directions,
say 1, we should obtain the IIB (9,1) algebra with the single
time direction in the 2 direction. Performing the
T-duality
\begin{equation}
Q^+ = Q^1\, ,\hskip 2truecm Q^- = \rmi \Gamma^1Q^2\, ,
\end{equation}
we indeed obtain the correct reality conditions for the
IIB (9,1) supergenerators,
see (\ref{Qstar}), with the matrix $\alpha^{(B)}_{9,1}$ given by
\begin{equation}
\alpha^{(B)}_{9,1} = -\rmi\, \alpha_{8,2}\, \tau_1\, .
\end{equation}
Another possibility is to T-dualize the IIA (8,2) algebra in one of the
8 spacelike directions, say 3. This duality proceeds similar to the
T-duality of the IIA (10,0) algebra discussed before and leads us to
the IIB (7,3) algebra with the 3 timelike directions in the 1,2 and 3
directions. Performing the T-duality
\begin{equation}
Q^+ = Q^1\, ,\hskip 2truecm Q^- = \rmi \Gamma^3Q^2\, ,
\end{equation}
we indeed obtain the reality conditions of the IIB (7,3) algebra,
see (\ref{symplMajIIB}), with
\begin{equation}
\alpha_{7,3} =- \rmi\, \alpha_{8,2}\, .
\end{equation}

Finally, one may also T-dualize in two different ways the IIA (6,4)
superalgebra. The reality conditions
on the supergenerators for this case are given by
\begin{equation}
{-1\over \alpha_{6,4}}\Gamma_1\Gamma_2\Gamma_3\Gamma_4
 C^{-1} (Q^+)^* = Q^-\, ,\hskip 2truecm
{-1\over \alpha_{6,4}}\Gamma_1\Gamma_2\Gamma_3\Gamma_4
C^{-1} (Q^-)^* = \,- Q^+\,,
\end{equation}
where 1,2,3,4 refer to the four timelike directions.
The first possibility is to T-dualize in one of the 6 spacelike directions,
say 5. This leads us to the IIB (5,5) algebra in exactly the same way
as the T-duality of the IIA (10,0) algebra in a spacelike direction
gave us the IIB (9,1) algebra so we will not discuss this case further.
The second possibility is to T-dualize one of the 4
timelike directions, say 1. This leads us to the IIB (7,3) algebra
with the 3 timelike directions in the directions 2,3 and 4.
Indeed, applying the T-duality rule
\begin{equation}
Q^+ = Q^1\, ,\hskip 2truecm Q^- = \rmi \Gamma^1Q^2\, ,
\end{equation}
we find the proper reality condition on the IIB (7,3) supergenerators
with
\begin{equation}
\alpha_{7,3} =- \rmi\,\alpha_{6,4}\, .
\end{equation}
This concludes our discussion of the real T-dualities. We have
summarized the situation in Table~\ref{ff:small}.

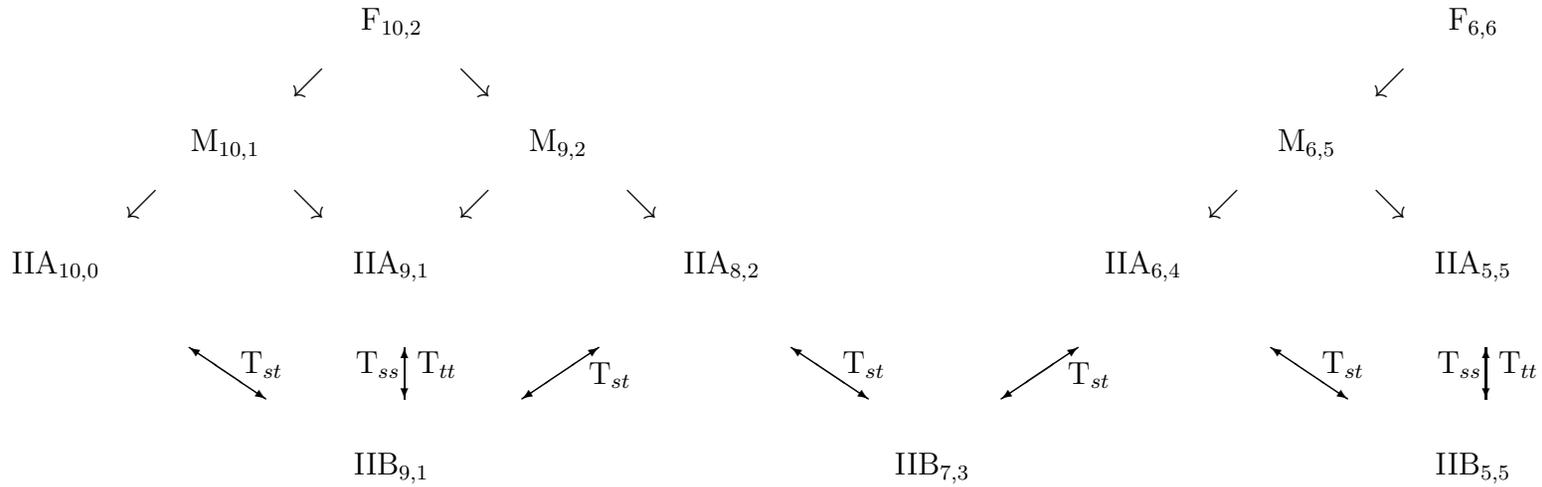
\landscape\tabcolsep 1pt
\begin{table}[ht]
\[
\begin{array}{cccccccccccccccccccc}
  &  &  &          &      \textrm{F}_{10,2}          &           &  &  &  &  &  &  &  &  &  &  &   \textrm{F}_{6,6} \\[3mm]
  &  &  &           \swarrow &   &\searrow              &  &  &  &  &  &  &  &  &  &   \swarrow  &  \\[3mm]
            &     &  \textrm{M}_{10,1} &                  &   &       & \textrm{M}_{9,2} &     &&  &  &  &  &  &    \textrm{M}_{6,5}&  &   \\[3mm]
            &\swarrow &     &\searrow&                 &  \swarrow&       &\searrow &    &  &  &  &  & \swarrow &    & \searrow  &    &  &  &  \\[3mm]
\textrm{IIA}_{10,0}  &   &  &  & \textrm{IIA}_{9,1}&&  &  & \textrm{IIA}_{8,2} &  &  &  & \textrm{IIA}_{6,4} &  &  &  & \textrm{IIA}_{5,5} &  \\[5mm]
            &   &  \nwsearrow     &  & \nsarrow  &&   \neswarrow      & &  &  \nwsearrow       &  &  \neswarrow &  &  & \nwsearrow   &  & \nsarrow  \\[5mm]
            &   &                 &  & \textrm{IIB}_{9,1} &  &  &  &  &  & \textrm{IIB}_{7,3} &    &  &  & & & \textrm{IIB}_{5,5}    \\[3mm]
\end{array}
\]
\caption{The realizations in $d=12$, 11 and 10 of the real algebra
   $OSp(1|32)$ for different signatures. The diagonal one-sided arrows indicate
dimensional reductions. The  both-sided arrows refer to the
space/space ($\rm T_{ss}$), time/time ($\rm T_{tt}$) and space/time ($\rm T_{st}$)
dualities discussed in the text.
\label{ff:small}   }
\end{table}\tabcolsep 6pt

\endlandscape

\section{The $*$-algebras}\label{ss:SUGRA}
In this section we will indicate
the connection between the algebras and spacetime for the different
theories. In this way, we will find a place for the $*$-algebras of
\cite{Hull}.

Up to here, all the generators that occurred at the right-hand side
of the $\{Q,Q\}$ anticommutators have been treated on an equal
footing. Now, we will identify two special operators.
\begin{description}
  \item[The spacetime translations] $P_\mu $. For associating the
  algebra with a field theory, one has to define spacetime points $x^\mu$.
  The spacetime is defined as the space generated from a basic point by
  these translation operators, as in a coset approach. In subsection~\ref{ss:translation}
 we will show how different choices of this translation generator
leads to the different $\star$-algebras of \cite{Hull}. The duality
relations between these $\star$-algebras is discussed in subsection~\ref{ss:Tdual*}.
  \item[The energy operator] $E$. This is the unique, as we will show below,
   operator
that is positive. It will be treated in subsection~\ref{ss:energy}.
\end{description}

\subsection{The translation operator}\label{ss:translation}
We first discuss the translation operator. It has to be
a vector in spacetime, and we will further only require that it
appears in a non-singular way\footnote{To be precise, we require that
the translation vector part of the matrix ${\cal Z}^{ik}$
as defined in (\ref{QQCZ}) is a non-singular matrix.}
 in the supersymmetry anticommutator.
This requirement is imposed in order that all supersymmetries act on
the spacetime.

There exists no vector operator in the 12-dimensional algebra, and
thus there is no translation operator in this case.
Indeed, for F theory there is no associated field theory. In 11
dimensions there is just one vector in the algebra (\ref{calg11}), and we thus
identify
\begin{equation}
  \tilde P_{\tilde M}\equiv  \tilde Z_{\tilde M}\,.
\label{defP11}
\end{equation}
The situation becomes more interesting in 10 dimensions, where there
are several vectors. In the IIA theory (\ref{calgIIA}) there are 2
vectors, $Z_M^+$ and $Z_M^-$. Each separately is not suitable as a
translation generator as e.g. $Z_M^+$ does not appear in the
anticommutator of the $Q^-$ supersymmetries. Instead, we may
consider any linear
combination of these two as a possible translation generator:
\begin{equation}
  P_M= p_+ Z_M^+ +p_- Z_M^-\,.
\label{PMIIA}
\end{equation}
Different values of $p_+$ and $p_-$ can be related by redefinitions
of the supersymmetries. Redefining $Q^\pm $ by a factor $q^\pm $, the
$p_\pm $ are redefined to $p_\pm q_\pm ^2$.

When going from complex to real algebras, there is a further restriction,
namely we want a
real\footnote{We will use $P^\ncc=P$ as reality condition, as the
difference with $P^*=P$ are just phase factors.} translation
operator.
For the (9,1) signature the reality condition on $P_M$ and $Z_M^\pm $
imply that $p_\pm$ are real. On the other hand, the reality
conditions on the supersymmetries imply that $q_\pm $ are real, for
fixed reality conditions on the supersymmetries, i.e. fixed matrix
$\alpha _{9,1}^{(A)}$. We now consider the basis where $\alpha
_{9,1}^{(A)}=\unity $.
Then, up to redefinitions and an overall sign, there are 2 possible choices for
the translation generator:
\begin{eqnarray}
(9,1)\ :\ {\rm IIA}\phantom{*} & : & P_M\equiv Z_M^+ + Z_M^- \nonumber\\
{\rm IIA}^* & : &P_M\equiv Z_M^+ - Z_M^-\,.
\label{defIIA*}
\end{eqnarray}
In Lagrangian theories, the spacetime is defined, and thus also the
translation operators are defined. In the IIA theories, they appear
in the algebra as the operator in the first line of (\ref{defIIA*}).
On the other hand, in the so-called IIA* theories of \cite{Hull},
translations appear as in the second line of (\ref{defIIA*}). The
complete algebras of the two theories only differ in the
identification of the translation operator.

In the case of the
(10,0) or (8,2) IIA algebras, the reality condition of $Z_M^\pm $
in (\ref{hermZE}) or (\ref{hermZ82}) implies $p_-=-p_+^*$, and the
reality conditions on the supersymmetries allow redefinitions with
$q_-=q_+^*$. These are sufficient to transform the translation operator to
the one with $p_+=p_-=1$. In these cases there is thus only one
translation operator,
\begin{equation}
 \textrm{IIA}\ : (10,0)\mbox{ or }(8,2)\ :\ P_M\equiv Z_M^+ + Z_M^-\,.
\label{defP100}
\end{equation}
Of course, we could also have chosen $p_+=\rmi$, in which case
\begin{equation}
  \textrm{IIA}'\ : (10,0)\mbox{ or }(8,2)\ :\ P_M\equiv \rmi(Z_M^+ -
  Z_M^-)\,.
\label{defPIIAp}
\end{equation}
This choice, which we label by $\rm {IIA}'$, is thus equivalent to
the choice labelled IIA, by a redefinition of the generators. This
redefinition is similar to the dualities treated at the end of
subsection~\ref{ss:calg1210}.
The obtained results for the IIA algebras
can of course be generalised from $(t,s)$ to
$(t+4,s-4)$ signatures.

The possibilities for choosing the translation operator in the IIB
algebras follow a similar pattern. For (9,1) (or (5,5)) signature,
there are two inequivalent choices of the translation operator: in
the basis where $\alpha =\unity $ these can be
\begin{eqnarray}
(9,1)\ :\hskip .2truecm  &&\hskip .4truecm {\rm IIB}\phantom{*} \ : \
P_M=Y^{(0)}_M\, , \nonumber\\
&&\left\{
\begin{array}{l}
{\rm IIB}^* \  : \ P_M=Y^{(3)}_M\, , \\
{\rm IIB}' \  : \ P_M=Y^{(1)}_M\, .
\end{array}
\right.
\label{PIIB}
\end{eqnarray}
The two choices ${\rm IIB}^*$ and $\rm {IIB}'$, following the notations of \cite{Hull},
are equivalent up to a redefinition,
which is the $S$-duality transformation given in (\ref{S}).

We stress that, in IIB as well as in IIA theories, the adapted
terminology refers to a basis in which $\alpha =\unity $. If $\alpha
$ is a different matrix, then one first has to redefine the
supersymmetries with $\alpha ^{1/2}$:
\begin{equation}
  Q'^i=\left( \alpha ^{1/2}\right) ^i{}_j Q^j\, .
\label{Q'}
\end{equation}
The newly defined generators $Q'^i$ satisfy reality conditions with
$\alpha '=\unity $. The
bosonic operators appearing in the anticommutator
\begin{equation}
  \left\{ Q^i,Q^j\right\} ={\cal Y}^{ij}
\label{QQcY}
\end{equation}
are in the new basis given by
\begin{equation}
   {\cal Y}'^{ij}=\left( \alpha ^{1/2}\right) ^i{}_k {\cal Y}^{k\ell }
   \left( \alpha ^{1/2}\right) ^j{}_\ell\,,
\label{Y'}
\end{equation}
and these are to be compared with (\ref{defIIA*})--(\ref{PIIB}).

For the (7,3) signature, there is only one inequivalent
choice, similar to the (10,0) or (8,2) case in IIA:
\begin{equation}
  (7,3)\ :\ P_M=Y_{M}^{(0)}\,.
\label{P73}
\end{equation}
Another way to express this fact is to say that there are three
 S-dual versions.
\subsection{T dualities}\label{ss:Tdual*}

\begin{table}[ht]
\[
\begin{array}{cccccc}
    &\phantom{\longleftrightarrow}\hspace{2cm}  &\textrm{ M}_{10,1} & \phantom{\longleftrightarrow}   & \textrm{M}_{9,2} &
    \\[4mm]
    &   & \downarrow &   & \downarrow &   \\[4mm]
    &   & \textrm{IIA}_{9,1} &   &   \textrm{IIA}^*_{9,1} &   \\
    &   & \nsarrowss & \crosstt & \nsarrowss &   \\
  \multicolumn{5}{c}{\moeilijk} & \weg  \\
\end{array} \]
  \caption{Detail of Table~\ref{ff:small}, indicating various
  theories
  distinguished by the choice of the translation operator.
 Our notation is explained in the caption of Table~\ref{ff:small}. For a discussion
of this table, see subsection~\ref{ss:Tdual*}.}\label{ff:detail}
\end{table}

The theories distinguished in the previous section are represented in
Table~\ref{ff:detail}, which is in fact, a close look to part of
Table~\ref{ff:small}. These theories are connected by various
dualities, which we now discuss.

Let us start with the reduction from 11 dimensions, where the
translation operator is always (\ref{defP11}), or in 12-dimensional
language $\hat{Z}_{M\,12}$. The reduction to any 10-dimensional
theory always leads to a translation generator of the form
$Z_M^++Z_M^-$. However, the $\alpha $ matrix is not always $\unity $.
Indeed, from (\ref{resultalpha10}), we see that reducing via the
(10,1) theory gives $\alpha ^{(A)}_{9,1}$ proportional to the unit
matrix, but reducing via (9,2) gives instead a matrix proportional to
$\tau _3$. Therefore in the first case the translation operator is
the one for the IIA theory, while in the second case, we have to
apply the correction (\ref{Y'}), after which the translation operator
is $\rmi(Z_M^+-Z_M^-)$, thus the one of $\rm {IIA}^*$. This explains the
two vertical arrows at the top of Table~\ref{ff:detail}.

For \textrm{T} dualities in 10 dimensions, the translation generators of one algebra
are split in 1 component in the direction of the
duality direction and 9 components orthogonal to that direction. They
are mapped to components of different Lorentz-covariant vectors or
tensors in the dual algebra. To identify dual algebras, we consider
only how the 9 translations orthogonal to the duality direction
transform.

Under space/space dualities, the operators are related by
(\ref{Tdual}), and the $\alpha $ matrices remain in standard form,
following (\ref{alphaB91}). It follows that under space/space duality
IIA is mapped to IIB and $\rm {IIA}^*$ to $\rm {IIB}^*$. This is
indicated by the two vertical double-sided arrows in the middle
of Table~\ref{ff:detail}. Under time/time dualities, the
bosonic operators are related in the same way as in the previous case
(according to (\ref{Tdualtt}). However, a unit $\alpha^{(A)} $ matrix
is, according to (\ref{alphatt}), mapped to a $\tau _3$ matrix. Therefore,
$Y^{(0)}$ and $Y^{(3)}$ are interchanged by the transformation
(\ref{Y'}), resulting in the $\mathrm{T}_{tt}$ rules indicated in the
Table. Finally, under space/time dualities, the maps of the generators in
(\ref{Tdual10,0}) are the same, but the $\alpha^{(B)}_{9,1} $ matrix
is proportional to $\tau _1$ according to (\ref{alpha91100}).
When we start from IIA in (10,0), the translation generator is
proportional to the unit matrix in $(ij)$ space, which is, by the
mapping (\ref{Y'}), transformed to $\tau ^1$. Therefore, it is mapped
to the $\rm {IIB}'$ translation generator. On the other hand, in
the $\rm {IIA}'$ theory, the translation generator is proportional to $\tau
_3$, which is invariant under the redefinitions. Thus $\rm {IIA}'$ is
mapped to $\rm {IIB}^*$. These two space/time duality relations are
indicated by the lower two double-sided arrows in  Table~\ref{ff:detail}.

\subsection{The energy operator}\label{ss:energy}
We will now clarify the uniqueness of the energy operator\footnote{Of
course, one can redefine the energy operator by adding some other
charges and still obtain a positive definite operator. However, then the
amount of mixture of the other charges is limited, and our energy operator
still has to appear in the combination.}. A first
technical issue is the positivity of states generated by
supersymmetry creation operators. Remark that $\{Q^i, Q^{j\,*}\}$ can not be
positive in case $\beta =-1$, as its hermitian conjugate is $\beta
\{Q^{j\,*}, Q^i\}=\beta \{ Q^i,Q^{j\,*}\}$, and is thus not even real for $\beta
=-1$. This indicates that the positivity for fermion bilinears
is not a trivial issue. We can define the positivity for our purposes
by taking for a square root of $\beta $
\begin{equation}
  \sqrt{\beta }\left\{Q^i, Q^{j\,*}\right\} =
   \sqrt{\beta }\left\{Q^i, Q^{k\,C}\right\} \alpha^j{}_k  B^T \geq
   0\,.
\label{positiveQQ}
\end{equation}
Adopting the notation
\begin{equation}
  \left\{Q^i, Q^{k\,C}\right\}= \mathcal{Z}^{ik} {\cal C}^{-1}\,,
\label{QQCZ}
\end{equation}
where we have now written the charge conjugation explicitly,  we
obtain the positivity condition
\begin{equation}
 \sqrt{\beta }\mathcal{Z}^{ik}\alpha^j{}_k\Gamma ^t \cdots \Gamma ^1
\geq 0\,,
  \label{Zpos}
\end{equation}
where the timelike gamma matrices appear. Splitting ${\cal Z}$ in $
\Gamma^{(\Lambda )} Z_\Lambda $ where $\Gamma ^{(\Lambda )}$ are all the basic elements of the
Clifford algebra, we find that the operator connected to $\Gamma ^{1\cdots
t}$ contains the one that always has to be non-zero and positive. We
further have to make a projection in the $(ij)$ space. Let us show
this by an example: the IIB algebra in (9,1) spacetime. For Minkowskian
signature \eqs (\ref{Zpos}) contains only the timelike $\Gamma $ matrix,
say $\Gamma
^0$. The only operators that we thus have to consider in
(\ref{calgIIB}) is $Y_0^{ij}$. The condition (\ref{Zpos}) for states
with all other central charges  zero reads
\begin{equation}
\sqrt{\beta } Y_0^{ik}\alpha^j{}_k >0\,.
\label{posYij}
\end{equation}
For instance, for $\beta =1$ and $\alpha =\unity $, the operator
$Y_0^{(0)}$ in (\ref{Y013}) is thus the one that has to be positive,
and is identified as the energy operator. Without imposing
$\beta =1$ and $\alpha =\unity $, the reality condition on $Y_0^{ij}$
is (\ref{Ystar}), and one can check the consistency that the positive combination
(\ref{posYij}) is then indeed also real due to $\alpha\alpha ^* =\unity
$ and $(\sqrt{\beta })^*\beta =\sqrt{\beta }$.

Another illustrative example is provided by the case that the $\alpha$
matrix is given by (\ref{alpha91100}). This case appears
in the duality between the IIB (9,1) and IIA (10,0)
algebra. We now find from (\ref{posYij}) that
\begin{equation}
  -\rmi \sqrt{\beta } Y_0^{(1)}\alpha _{10,0}\geq 0\,.
\label{posYTdual}
\end{equation}
One checks again from (\ref{reality9,1}) that this energy operator is
real. Under space/time duality this operator is mapped onto
the operator $- \sqrt{\beta }Z \alpha
_{10,0}$ of the Euclidean IIA superalgebra.
We thus conclude that the energy operator of the Euclidean IIA algebra
is proportional to the scalar $Z$ generator.
This is consistent with the general rule (\ref{Zpos}) in case of an Euclidean
space. We conclude from this example that the energy operator
does not always occur as a component of the translation generator.

We found that the energy operator is a bosonic operator that appears in
the anticommutator of two supersymmetries with the product of all timelike
$\Gamma$ matrices $\Gamma ^{1\cdots t}$. An immediate consequence of
this, is that it can only be one of the generators of translations if
the space is Minkowskian. Indeed, the translations appear by
definition with one $\Gamma $ matrix in this anticommutator. In dimension
(10,1), this is the usual situation. The time component of the
translations is the energy operator.
But this is thus not the case in the (9,2) theory, where the energy
is part of the two-index central charge operator.

In non-Minkowskian signatures, energy is not part of the translations.
In Euclidean spacetime, it is part of a scalar central
charge. If there are 2 or more timelike
directions, it is part of a tensor central charge.

Finally, for the (9,1) theories, we can identify the energy operator
as being the timelike component of the vector operator that is identified
as the translation in the IIA and IIB theories. Thus, the
timelike component of the translation operators in IIA${}^*$,
IIB${}^*$ or IIB${}'$ algebras is \emph{not} the positive definite
operator. In the corresponding field theories, the terms with a timelike
derivative are often called `kinetic-energy terms'. As this time coordinate
corresponds to the timelike component of the translation generator,
this is not what corresponds to the positive definite `energy'
operator that we defined here. Therefore, the terms
that appear with derivatives $\partial _t$ in the
action, are not positive definite. This is indeed the characteristic
feature of the IIA${}^*$, IIB${}^*$ or IIB${}'$ Lagrangian theories.

\section{Conclusions}\label{ss:conclusions}

In this paper we have studied how the symmetries of F-theories,
M-theories and type II string theories exhibit many faces of
the $OSp(1|32)$ algebra, rewriting it covariantly in
dimensions
12, 11 and 10 of different signatures. One only has to
distinguish the complex algebra and the unique real algebra.
We have explicitly shown how,
both in the complex and in the real case, the $d=10$, 11 and 12
algebras are related by dimensional reduction. Furthermore, we have
shown how the (complex and real) $d=10$ IIA and IIB superalgebras are related
by T-duality (see Table~\ref{ff:small}). The identification of the translation generator
in these algebras leads to the $\star$-algebras of \cite{Hull}. This leads to
a more detailed set of T-dualities whose structure is indicated in Table~\ref{ff:detail}.

{}From a physical point of view one might prefer the Minkowskian
algebras above the other signatures since a priori it is not clear
how to make sense out of non-Minkowskian signatures. On the other hand, we have
seen that the different signatures are related to each other by a
generalized space/time T-duality map. One of the lessons we have
learned from the modern developments on dualities in string theory is that
what looks strange in one picture can perfectly make sense in the
dual picture. Many-time physics could make sense within the
context of an underlying string theory.
This was the philosophy adapted in \cite{Hull}.

There are several ways in which string or M-theory hints at the relevance
of non-Minkowskian signatures. For instance, they naturally occur in
a classification of supermembranes \cite{Blencowe:1988sk}. Also,
it has been suggested that string theory has
a twelve-dimensional F-theory origin with signature (10,2), i.e.~two times
\cite{Vafa:1996xn}.
For recent work on two-time physics, see \cite{Bars:2000mz} and
references therein. Also for embeddings of branes in flat spaces one
needs two time directions \cite{embedding}. Two times also occur when one considers N=2
superstrings leading to a $d=4$ target spacetime of (2,2) signature
\cite{Ooguri:1990ww}.
It would be interesting to see whether the superalgebraic approach
of this work could be applied to this case. Space/time T-duality might
then relate physics in (2,2) dimensions to the Minkowskian (3,1) signature.

Let us now return to the original motivation of this paper,
i.e.~the study of Type IIB string theory in (10,0) Euclidean dimensions. From
Table~\ref{ff:small} we see that the (10,0) signature only allows a real Euclidean
Type IIA theory. Therefore we confirm that, in order to discuss
the IIB case, one needs complexification,
as it has been the case for other Euclidean theories \cite{Nie96}.
Note that, after complexification, the theory does not refer to any
specific signature. The original D-instanton must be considered
in the context of this complex IIB theory.

At a first stage one can consider the complex T-dual of the D-instanton.
Under this map the D-instanton will be mapped to a T-dual two-block
solution of
a complex IIA supergravity theory. At a second stage, one can
impose reality conditions on the complex fields of the
IIA supergravity theory. These reality conditions not only specify
the signature but also, since the signs of the
kinetic terms in the action are now fixed, the choice of the
translation generator. The reality conditions thus lead to
a Minkowskian supergravity theory based on either the
${\rm IIA}_{9,1}$- or the ${\rm IIA}_{9,1}^\star$-algebra.
Since the D-instanton solution itself is real, we expect that
the dual solution can be embedded either in the  ${\rm IIA}_{9,1}$-
or the ${\rm IIA}_{9,1}^\star$-supergravity theory. The first
embedding corresponds to the standard D0-brane solution of a
 ${\rm IIA}_{9,1}$-supergravity theory as discussed in the introduction.
The second T-dual picture
of the D-instanton is, in the terminology of \cite{Hull},
an E1-brane of a  ${\rm IIA}_{9,1}^\star$-supergravity
theory.

It would be interesting to verify the existence of these two different
T-dual pictures of the D-instanton by constructing explicitly
the corresponding supergravity theories and T-duality maps.

\medskip
\section*{Acknowledgments}

\noindent
We are grateful to D.~Berman, S.~Cucu, G.~Gibbons, M.~G\"{u}naydin,
P.~Howe, C.M.~Hull, S. Vandoren, P.~van Nieuwenhuizen,
J.P.~van der Schaar and P.~Sundell for interesting and very useful
discussions.
This work was stimulated by discussions with Peter van Nieuwenhuizen at
the 1998 Amsterdam Summer Workshop where the issue of complexification
of selfdual
tensors in 6 Euclidean dimensions was discussed.
This work was
supported by the European Commission TMR programme
ERBFMRX-CT96-0045, in which E.B. is associated
to Utrecht University.
\newpage
\appendix
\section{Conventions}\label{app:conventions}
The conventions of this paper can be found in
\cite{VanProeyen:1999ni}. In any
even dimension $d$, with $t$ timelike directions and $s$ spacelike
dimensions, we define a chiral projection as follows:
\begin{equation}
  \Gamma _*= (-\rmi)^{(s-t)/2}\ft1{d!}\varepsilon _{a_1\cdots a_d}\Gamma ^{a_1\cdots a_d}
 \,,\qquad  {\cal P}^{\pm }=\ft12(\unity \pm \Gamma _*)\,.
\label{defGamma*}
\end{equation}
The dual of a rank $n$
antisymmetric tensor in $d=2n$ dimensions is defined by
\begin{equation}
  \tilde F_{a_1\ldots a_n}= (\rmi)^{d/2+t} \frac{1}{n!}
  \varepsilon _{a_1\ldots a_d}F^{a_d\ldots a_{n+1}}\,,
\label{dualF}
\end{equation}
such that
\begin{eqnarray}
 && \tilde {\tilde F}_{a_1\ldots a_n} =F_{a_1\ldots a_n}\,,\qquad
\Gamma _{a_1\ldots a_n}=\Gamma _*\tilde \Gamma _{a_1\ldots a_n}\, , \nonumber\\
&& {\tilde F}_{a_1\ldots a_n}{\tilde G}^{a_1\ldots
a_n}=(-)^n F_{a_1\ldots a_n}G^{a_1\ldots a_n}
\,.
\label{dualrelations}
\end{eqnarray}
These are used to define (anti-)self-dual tensors
\begin{equation}
  F^\pm =\ft12(F\pm \tilde F)\,.
\label{Fpm}
\end{equation}

For the formulae of gamma matrices in dimensions ten to twelve,
we only need the cases where
$\epsilon =\eta =1$ in the terminology of \cite{VanProeyen:1999ni}.
For 10 and 12 dimensions, this is a choice, while for 11 dimensions
this is necessary. Therefore, for the purpose of this paper, we can suffice
with part of the general formalism.
 With these choices, the essential ingredients are the
following. One has a charge conjugation matrix $ {\cal C}$, which is
unitary and antisymmetric, and satisfies
the property
\begin{equation}
\Gamma _{a}^T  =  - {\cal C} \Gamma_a {\cal C}^{-1}\, .
\label{CT}
\end{equation}
Another unitary matrix that plays an important role
in the definition of complex conjugation of spinors (see below) is
\begin{equation}
  B=-{\cal C}\Gamma _1\cdots \Gamma _t\,,\qquad
  B^{-1}=(-)^{(t-1)(t-2)/2}\Gamma _1\cdots \Gamma _t{\cal C}^{-1}\,,
\label{defB}
\end{equation}
where $\Gamma _1,\cdots ,\Gamma _t$ are the gamma matrices in the $t$ timelike
directions. An important property is
\begin{equation}
  B^*\, B = (-)^{(t-1)(t-2)/2}\,.
\label{B*B}
\end{equation}

We do not write spinor indices in the main text, and neither the
charge conjugation matrices which should always appear at the
right-hand side of anticommutation relations. For instance, we write
\begin{equation}
  \left\{ Q_\alpha ,Q_\beta \right\} = (\Gamma ^\mu )_\alpha {}^\gamma
  ({\cal C}^{-1})_{\gamma \beta } P_\mu \hskip .5truecm
\longrightarrow \hskip .5truecm
  \left\{ Q,Q\right\} =\Gamma ^\mu P_\mu \,.
\label{shortanticomm}
\end{equation}
Using this shorthand notation $\Gamma _{a_1\cdots a_r}$ is symmetric for $r=1,2$ mod~4,
and antisymmetric for $r=0,3$ mod~4.

Complex conjugation can be conveniently performed by replacing it
with `charge conjugation'. This operation works on spinors as
\begin{equation}
  \lambda ^C=\frac{1}{\alpha } B^{-1} \lambda ^*\,,
\label{lambdaC}
\end{equation}
where $\alpha $ is an arbitrary phase factor. When there are
several spinors $\lambda ^i$, then $\alpha $ may even be a matrix
$\alpha ^i{}_j$, such that
\begin{equation}
  \alpha ^{*i}{}_j\alpha ^j{}_k=\delta ^i{}_k\,.
\label{alphaalpha*}
\end{equation}
On matrices in spinor space, $M$, charge conjugation acts as
\begin{equation}
  M^C= B^{-1}M^* B\, .
\label{MC}
\end{equation}
On spinors, the square of the $C$ operation is
determined by (\ref{B*B}), i.e.\ it is the identity for $t=1,2$
mod~4, and minus the identity for $t=0,3$ mod~4.
A Majorana spinor is a spinor that is invariant under this charge
conjugation, which is thus only possible\footnote{Note that
for $t=0$ mod 4, a Majorana spinor would be possible if we first
replace ${\cal C}$ by $\rmi {\cal C}\Gamma _*$. This would amount to
the other signs of parameters ($\epsilon =\eta =-1$) in
\cite{VanProeyen:1999ni}. However, in this paper we will always
choose $\epsilon=\eta = 1$. With this choice, we cannot define
Majorana spinors for $t=0,3$ mod~4, but we can define symplectic
Majorana spinors (see
Table~\ref{tbl:MWSspinors} in Section~\ref{ss:realalgebras}).
} for $t=1,2$ mod~4.

The charge conjugation operator applied to anticommutation relations,
amounts in practice to the following rules.
In the left-hand side,
replace the fermionic generators by their $C$ conjugate. In the
right-hand side, add a factor $\beta\alpha ^{-2} $ (if $\alpha $ is just
a number), and apply the $C$
operation to the gamma
matrices and generators (for which it is just hermitian conjugation).
The parameter $ \beta $ is a sign that depends on the convention whether
complex conjugations maintains ($\beta =1$) or interchanges ($\beta
=-1$) the order of fermions.
One can then neglect the unwritten charge conjugation matrix.
Thus e.g.\ applying it to (\ref{shortanticomm}), it gives
\begin{equation}
  \left\{ Q^C,Q^C\right\} =\alpha ^{-1} \beta (\Gamma ^\mu )^C P_\mu ^*
  \alpha ^{-1\,T}\, .
\label{QCQC}
\end{equation}
In this example we can commute $\alpha ^{-1\,T}$ through $P_\mu $,
but we write the expression in this way to indicate its form when
$P_\mu $ is replaced by a nontrivial matrix in the space of several
supersymmetries $Q^i$. Usually, $\alpha $ is just a number, and the
factors thus combine to $\beta \alpha ^{-2}$.

It is convenient to use a $\ncc$-notation for the bosonic generators
that hides the prefactor. We thus define for all bosonic generators $Z$
\begin{equation}
  Z^\ncc\equiv \beta\alpha ^{-1} Z^*\alpha ^{-1\,T} \, ,
\label{B*}
\end{equation}
e.g.~$P_\mu^\ncc = \beta\alpha^{-2} P_\mu^*$.
In practice the factor $\beta $ is a choice which is independent of
the spacetime dimension or signature. One could choose $\beta =1$. On
the other hand, the $\alpha $ factors vary for different spacetime
dimensions and signatures, and are thus relevant.

\section{Gamma matrices in 10 to 12 dimensions}
\label{app:gamma}

To distinguish the different gamma matrices in 10 to 12 Euclidean
dimensions we use the following notation:

\begin{equation}
\begin{array}{lll}
d=10:&\Gamma_M&M=1,\cdots ,10\, ,\\
d=11:&{\tilde {\Gamma}}_{\tilde M}&{\tilde M} = 1,\cdots ,10,11\, ,\\
d=12:&{\hat {\Gamma}}_{\hat M}& {\hat M} = 1,\cdots ,10,11,12\, .
\end{array}
\end{equation}
The gamma matrices for other spacetime signatures are obtained via
the relations
\begin{eqnarray}
\Gamma^t &=& -\rmi \Gamnul{}^t\, ,\nonumber\\
\Gamma^s& =& \Gamnul{}^s\,,
\end{eqnarray}
for any timelike direction $t$ and spacelike direction $s$.
Here $\Gamnul$ are the gamma matrices corresponding to
the Euclidean signature.
It is thus sufficient to construct the Euclidean gamma matrices
which, for simplicity, we denote in the following by $\Gamma$ instead
of $\Gamnul$.

Our starting point is the set of
$32\times 32$ matrices $\Gamma_M$ in 10 Euclidean dimensions
which are the basic building blocks of our construction.
Using the general definition (\ref{defGamma*}) for $s=10$, $t=0$, we have
\begin{equation}
\Gamma _*=-\rmi \Gamma _1\ldots \Gamma _{10} \equiv \Gamma_{11}\, .
\end{equation}
The $\Gamma_M$ together with $\Gamma_{11}$ define
the gamma matrices in 11 Euclidean dimensions.

To construct the gamma matrices in twelve Euclidean dimensions
we define the $64 \times 64$ matrices
\begin{eqnarray}
\hat{\Gamma }_M & = & \Gamma _M\otimes \unity_{32}\, ,  \nonumber\\
\hat{\Gamma }_{11} & = & \Gamma _{11}\otimes \sigma _1\, ,\nonumber\\
\hat{\Gamma }_{12} & = & \Gamma _{11}\otimes \sigma _2\, ,\nonumber\\
\hat{\Gamma }_{*} & = &-\hat{\Gamma }_1
\ldots \hat{\Gamma }_{12}= \Gamma _{11}\otimes \sigma _3\,.
\label{Gamma64}
\end{eqnarray}

All charge conjugation matrices we use in this paper are unitary
and antisymmetric, i.e.
\begin{equation}
\begin{array}{lll}
d=10,11:& C^\dagger C = \unity_{32}\, ,& C^T = -C\, ,\\
d=12:& {\hat C}^\dagger {\hat C} = \unity_{64}\, , &{\hat C}^T =
- {\hat C}\, .
\end{array}
\end{equation}
In the notation of
\cite{VanProeyen:1999ni}, we always use $C=C_+$ ($d=10,11$)
and ${\hat C} = {\hat C}_+$ ($d=12$), i.e.~$\eta = {\hat \eta} = 1$.
The charge conjugation matrix  $C$ of 10 dimensions can also be used as
charge conjugation in eleven dimensions.
The $64 \times 64$ $d=12$ charge conjugation matrix $\hat C$
is constructed as follows:
\begin{equation}
\hat{C}  =  C \otimes \sigma _1\, .
\label{C+-64}
\end{equation}
All these charge conjugation matrices satisfy the
property (\ref{CT}).


\end{document}